%% file: simon_ufds_v8.tex
\documentclass[a4paper]{ar-1col}

\jname{Xxxx. Xxx. Xxx. Xxx.}
\jvol{AA}
\jyear{YYYY}
\doi{10.1146/((please add article doi))}

\usepackage{natbib}
\usepackage{aas_macros}
\usepackage{amssymb}
\usepackage{color}
\usepackage{deluxetable}
\usepackage{longtable}
\usepackage{lscape}

\newcommand\phn{\phantom{0}}%
\newcommand\phs{\phantom{$-$}}%
\newcommand{\kms}         {km~s$^{-1}$}
\newcommand{\msun}        {M$_{\odot}$}
\newcommand{\lsun}        {L$_{\odot}$}
\newcommand{\feh}         {[Fe/H]}
\newcommand{\hi}          {H~{\sc i}}

\begin{document}

\markboth{Simon}{The Faintest Dwarf Galaxies}

\title{The Faintest Dwarf Galaxies}

\author{Joshua D. Simon$^1$
\affil{$^1$Observatories of the Carnegie Institution for Science,
  Pasadena, USA, 91101; email: jsimon@carnegiescience.edu}
}

\begin{abstract}
The lowest luminosity ($L < 10^{5}$~L$_{\odot}$) Milky Way satellite
galaxies represent the extreme lower limit of the galaxy luminosity
function.  These ultra-faint dwarfs are the oldest, most dark
matter-dominated, most metal-poor, and least chemically evolved
stellar systems known.  They therefore provide unique windows into the
formation of the first galaxies and the behavior of dark matter on
small scales.  In this review, we summarize the discovery of
ultra-faint dwarfs in the Sloan Digital Sky Survey in 2005, and the
subsequent observational and theoretical progress in understanding
their nature and origin.  We describe their stellar kinematics,
chemical abundance patterns, structural properties, stellar
populations, orbits, and luminosity function, and what can be learned
from each type of measurement.  We conclude that: (1) In most cases,
the stellar velocity dispersions of ultra-faint dwarfs are robust
against systematic uncertainties such as binary stars and foreground
contamination; (2) The chemical abundance patterns of stars in
ultra-faint dwarfs require two sources of r-process elements, one of
which can likely be attributed to neutron star mergers; (3) Even under
conservative assumptions, only a small fraction of ultra-faint dwarfs
may have suffered significant tidal stripping of their stellar
components; (4) Determining the properties of the faintest dwarfs out
to the virial radius of the Milky Way will require very large
investments of observing time with future telescopes.  Finally, we
offer a look forward at the observations that will be possible with
future facilities as the push toward a complete census of the Local
Group dwarf galaxy population continues.
\end{abstract}

\maketitle

\tableofcontents

\section{INTRODUCTION}
\label{sec:intro}

The search for faint dwarf galaxies has been a nearly continuous
endeavor since the serendipitous discovery of the first such system,
Sculptor, by \citet{shapley38}.  As significantly deeper survey data
became available, systematic searches for more dwarfs slowly revealed
what are now known as the classical dwarf spheroidal (dSph) satellites
of the Milky Way \citep{shapley38b,hw50,wilson55,cannon77}.  However,
after the identification of Sextans by \citet{irwin90}, the push to
ever lower luminosities and surface brightnesses stalled for more than
a decade.  New efforts to find faint, low surface brightness Milky Way
dwarf galaxies continued fruitlessly in this period
\citep{kleyna97,sb02,willman02,hopp03,whiting07}.  Notably, though,
there were strong theoretical reasons to expect that dwarfs with
substantially lower luminosities and surface brightnesses should exist
\citep{benson02}.

\begin{figure}[h]
\includegraphics[width=6.33in]{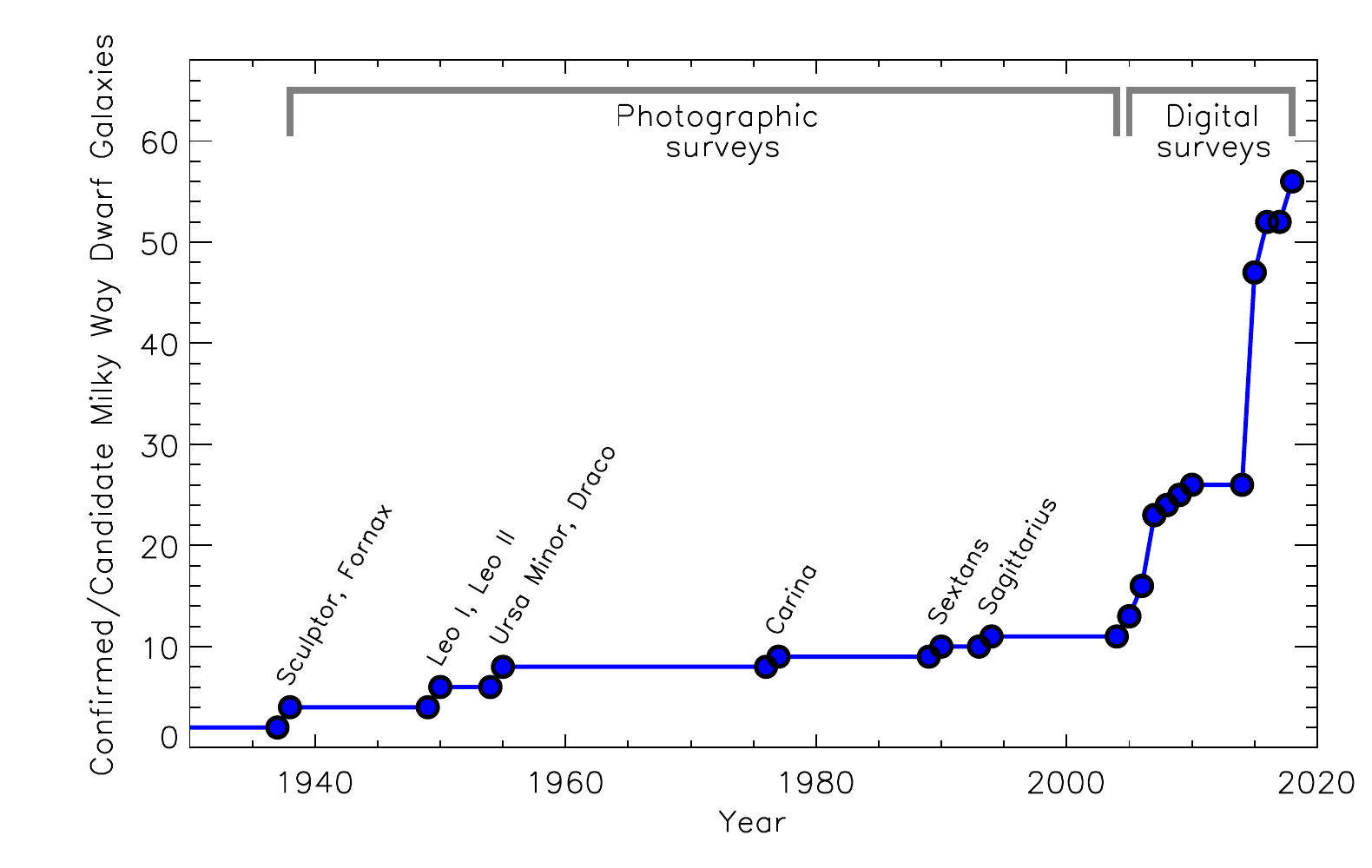}
\caption{Census of Milky Way satellite galaxies as a function of time.
  The objects shown here include all spectroscopically confirmed dwarf
  galaxies as well as those suspected to be dwarfs based on less
  conclusive spectroscopic and photometric measurements.  The major
  discovery impact of SDSS (from 2005-2010) and DES/Pan-STARRS (2015),
  each of which approximately doubled the previously known satellite
  population, stands out in this historical perspective.}
\label{fig:dwarfs_vs_time}
\end{figure}

This prediction proved resoundingly correct in 2005, when the first
such objects were discovered in Sloan Digital Sky Survey (SDSS)
imaging by \citet{willman05a,willman05b}.  These results opened the
floodgates, and within two years the known population of Milky Way
satellite galaxies more than doubled
\citep{zucker06a,zucker06b,belokurov06,belokurov07,sh06,irwin07,walsh07}.
Over the following decade, new discoveries continued at a rapid pace
in SDSS and other surveys
\citep[e.g.,][]{belokurov08,belokurov09,belokurov10,bechtol15,koposov15,koposov18,drlica15,dw16,martin15,kim15peg3,kj15,laevens15,laevens15b,torrealba16b,torrealba18,homma16,homma18},
such that the Milky Way satellite census has now doubled yet again
(Figure~\ref{fig:dwarfs_vs_time}).  Thanks to significant investments
of telescope time in deep imaging and spectroscopy of the newly
discovered objects, along with accompanying theoretical modeling, we
now have a general understanding of the properties of these systems
and their place in galaxy evolution and cosmology.

While the faintest dwarf galaxies resemble globular clusters in some
ways, when the population of low luminosity stellar systems is
considered as a whole it is clear that they are galaxies rather than
star clusters: (1) The stellar kinematics of ultra-faint dwarfs (UFDs)
demonstrate that they contain significant amounts of dark matter; (2)
All but the very lowest-luminosity UFDs have physical extents larger
than any known clusters; (3) Within each UFD, the abundances of Fe and
$\alpha$-elements exhibit substantial spreads resulting from extended
star formation and internal chemical enrichment; (4) UFDs follow a
luminosity-metallicity relationship, while globular clusters do not;
(5) The abundances of certain elements in UFDs are similar to those in
brighter dwarfs, and do not resemble the light element chemical
abundance correlations seen in globular clusters.  Each of these
results is discussed in more detail in the remainder of this article.

In this review we summarize the progress that has been made in
characterizing the least luminous galaxies since their discovery.  We
begin by motivating the study of the least luminous galaxies and by
offering a definition of the term ``ultra-faint dwarf,'' which has
been in common usage since the initial discoveries.  In
Section~\ref{sec:kin} we discuss the stellar kinematics and mass
modeling of UFDs, and the corresponding evidence that they are
galaxies rather than diffuse star clusters.  In
Section~\ref{sec:metallicity} we describe the metallicities and
chemical abundance patterns of stars in UFDs, including the
mass-metallicity relation, the chemical evolution of the smallest
dwarfs, and their role in establishing the site of r-process
nucleosynthesis.  We briefly summarize the structural properties of
the UFD population in Section~\ref{sec:struc}.  In
Section~\ref{sec:stellar_pops} we introduce the star formation
histories and initial mass functions of UFDs, and in
Section~\ref{sec:lf} we examine constraints on the luminosity function
of the faintest galaxies.  We consider the origin and evolution of
these systems based on theoretical work and measurements of their
orbits around the Milky Way in Section~\ref{sec:origin}.  We provide a
brief overview of the manifold ways in which UFDs may be used to
constrain the behavior of dark matter in Section~\ref{sec:dm}.  In
Section~\ref{sec:beyond_mw} we introduce the study of ultra-faint
dwarfs outside the immediate neighborhood of the Milky Way and the
connection between faint dwarfs in the Local Group and the
high-redshift universe.  In Section~\ref{sec:summary} we summarize the
current state of the field and suggest future paths for research.

\subsection{The Cosmological Significance of the Lowest Luminosity
  Dwarf Galaxies}

A reasonable astronomer might ask how the smallest, most inconspicuous
galaxies ever formed could have broad importance to the field of
astrophysics.  However, several aspects of the UFDs make them critical
objects to understand, with potentially wide-ranging implications.
First, UFDs reside in the smallest dark matter halos yet found.  While
only the mass at the very center of the halo is currently measurable,
the extrapolated virial masses of UFDs are
$\sim10^{9}$~\msun\ \citep[e.g.,][]{strigari08}, and the halo masses
at the time when the stars formed may have been
$\sim10^{8}$~\msun\ \citep[e.g.,][]{br09,safarzadeh18}.  UFDs are also
the most dark matter-dominated systems known.  This combination of
small halo mass and negligible baryonic mass makes UFDs extremely
valuable laboratories for constraining the nature of dark matter.
Simply counting the number of such objects around the Milky Way places
a limit on the mass of the dark matter particle
\citep[e.g.,][]{jethwa18}.  The census and observed mass function of
low-mass halos will point the direction toward solving the
long-standing and highly contentious missing satellite problem
\citep[e.g.,][]{klypin99,moore99,sg07,brooks13,kim18}.  The measured
central densities, and perhaps eventually the density profiles, of
UFDs provide significant clues to the behavior of dark matter on small
scales \citep[e.g.,][]{cs16,bozek18,errani18}.

Second, UFDs represent the extreme limit of the galaxy formation
process.  They have the lowest metallicities, oldest ages, smallest
sizes, smallest stellar masses, and simplest assembly histories of all
galaxies.  Both observations and theoretical models indicate that UFDs
formed at very high redshift, probably before the epoch of
reionization.  Unlike essentially all larger systems, they underwent
little to no further evolution after that time, and have survived to
the present day as pristine relics from the early universe
\citep[e.g.,][]{br09,br11a,wheeler15}.  These objects therefore
present us with a unique window into the conditions prevalent at the
time when the first galaxies were forming.

To our knowledge, no previous reviews have focused primarily or
exclusively on the properties of the faintest dwarf galaxies.
\citet{willman10} presented the first summary of searches for UFDs.
There have been many reviews on the broader population of dwarfs
\citep[e.g.,][the latter two of which also discuss
  UFDs]{mateo98,tolstoy09,mcconnachie12}, and various aspects of UFDs
have been featured in recent reviews on dark matter
\citep[e.g.,][]{bbk17,strigari18} and metal-poor stars
\citep[e.g.,][]{fn15}.  Given the rapid maturation of the study of the
very lowest luminosity galaxies over the last decade, here we aim to
provide a comprehensive discussion of the current state of knowledge
of these systems.  After first results from LSST become available,
some of this material may need to be revisited.

\subsection{Defining ``Ultra-Faint Dwarf''}
\label{sec:definition}

The dwarf galaxies known prior to 2005 have absolute magnitudes
brighter than $M_{{\rm V}} = -8.7$, corresponding to V-band
luminosities larger than $2.5 \times 10^{5}$~L$_{\odot}$.  Their
Plummer (half-light) radii are $\gtrsim200$~pc, and with the exception
of Sextans and Ursa Minor, their central surface brightnesses are
$<26$~mag~arcsec$^{-2}$.  In contrast, the dwarfs discovered in SDSS
and other modern surveys are up to a factor of $\sim1000$ less
luminous, with half-light radii as small as $\sim20$~pc and surface
brightnesses that can be $\sim2-3$~mag~arcsec$^{-2}$ fainter than that
of Sextans.

As was evident even from the titles of some of the first SDSS
discovery papers --- e.g., ``A New Milky Way Companion: Unusual
Globular Cluster or Extreme Dwarf Satellite?''\footnote{Indeed, the
  classification of Willman~1 is still not entirely secure, although
  the metallicities of its brightest member stars suggest that it is,
  or was, a dwarf galaxy \citep{willman11}.} \citep{willman05a} and
``A Curious Milky Way Satellite in Ursa Major''\footnote{Although
  \citet{zucker06b} argued that Ursa~Major~II is a dwarf galaxy, the
  same system was identified independently by \citet{grillmair06}, who
  described it as ``a new globular cluster or dwarf spheroidal.''}
\citep{zucker06b} --- the nature of these new satellites was not
immediately clear.  Over the next several years, spectroscopy of stars
in these objects pointed strongly to the conclusion that they were
dwarf galaxies rather than globular clusters
\citep{kleyna05,munoz06,martin07,sg07}.  Given the clear differences
in global properties relative to previously known dwarf galaxies, the
community rapidly began referring to these objects as ``ultra-faint''
dwarfs, a term first used by \citet{willman05a}.  However, no formal
definition of such a class was ever offered in the literature, and the
usage of it has not been entirely consistent.  In particular,
Canes~Venatici~I (CVn~I) is often referred to as a UFD because it was
discovered in SDSS data around the same time as many fainter dwarfs
\citep{zucker06a}, but its size and luminosity are nearly identical to
those of Ursa Minor, which was identified more than 50 years earlier
thanks to its location $\sim3\times$ closer to the Milky Way.

\begin{figure}[h]
\includegraphics[width=6.33in]{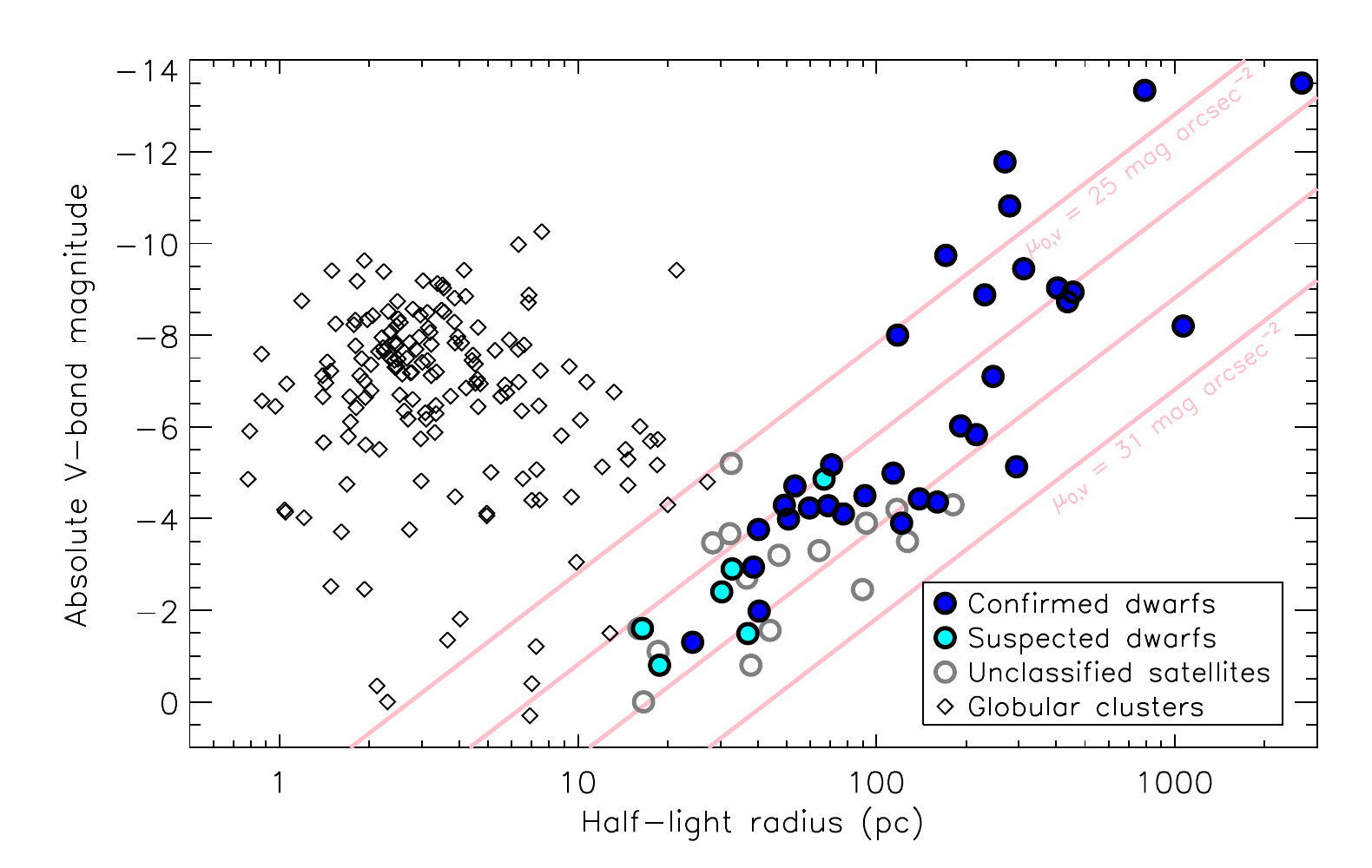}
\caption{Distribution of Milky Way satellites in absolute magnitude
  ($M_{{\rm V}}$) and half-light radius.  Confirmed dwarf galaxies are
  displayed as dark blue filled circles, and objects suspected to be
  dwarf galaxies but for which the available data are not conclusive
  are shown as cyan filled circles.  Dwarf galaxy candidates without
  any published classification (usually because of the lack of
  spectroscopy) are shown as open gray circles.  The faint candidates
  with $R_{1/2} \gtrsim 50$~pc are almost certainly dwarf galaxies,
  but we do not include them in the confirmed category here given the
  currently available observations.  The dwarf galaxy/candidate data
  included in this plot are listed in Table 1.  The black diamonds
  represent the Milky Way's globular clusters \citep{harris96}.  Lines
  of constant central surface brightness (at 25, 27, 29, and 31
  mag~arcsec$^{-2}$ in V band) are plotted in pink.  For stellar
  systems brighter than $M_{{\rm V}} \approx -5$ there is no ambiguity
  in classification: globular clusters have $R_{1/2} \lesssim 20$~pc
  and dwarf galaxies have $R_{1/2} \gtrsim 100$~pc.  At fainter
  magnitudes the size distributions begin to impinge upon each other
  and classification based purely on photometric parameters may not
  always be possible.  Whether the two populations actually occupy
  non-overlapping portions of this parameter space remains to be
  determined from spectroscopy of the faintest stellar systems with
  half-light radii between 10 and 40~pc.}
\label{fig:mv_rhalf}
\end{figure}

Despite this new nomenclature, it is now obvious that ultra-faint
dwarfs continuously extend the properties of more luminous dwarfs in
stellar mass, surface brightness, size, dynamical mass, and
metallicity (see Figure~\ref{fig:mv_rhalf} and
Sections~\ref{sec:kin}-\ref{sec:struc}).  They are not a physically
distinct class of objects.  Nevertheless, there are several reasons
why it may be useful to refer to them via a separate label.  In
particular, UFDs represent the extreme end (we presume) of the
distribution of galaxy properties, orders of magnitude beyond the
previously-known dwarfs in some respects.  Moreover, while classical
dSphs can already be identified and studied in other nearby groups of
galaxies, the UFDs are special in that only the very brightest
examples of such systems will be detectable beyond the Local Group in
the foreseeable future.  Because of their lack of bright stars,
detailed spectroscopic characterization of ultra-faint dwarfs will
likely remain limited to satellites of the Milky Way.  Finally, it is
tempting to suggest that UFDs might differ from classical dSphs in
that their star formation was shut off by reionization at $z \gtrsim
6$ instead of continuing to lower redshift.  While this hypothesis is
consistent with the available data, the sample of $M_{{\rm V}} \gtrsim
-9$ dwarf galaxies with precision star formation histories is too
small to draw firm conclusions yet.  If this idea turns out to be
correct, it would provide a physically-motivated division between
ultra-faint and classical dwarfs.

Based on the above discussion, we suggest that dwarf galaxies with
absolute magnitudes fainter than $M_{{\rm V}} = -7.7$ ($L =
10^{5}$~L$_{\odot}$) be considered UFDs.  This definition matches the
naming convention adopted by \citet{bbk17}.  Among the post-2005
discoveries, only four galaxies are within 1~magnitude of this
boundary: CVn~I ($M_{{\rm V}} = -8.7$), Crater~II ($M_{{\rm V}} =
-8.2$), Leo~T ($M_{{\rm V}} = -8.0$), and Eridanus~II ($M_{{\rm V}} =
-7.2$).  The first three of these systems stand out from the fainter
population in obvious ways: CVn~I is substantially more luminous,
larger, and more metal-rich
\citep[e.g.,][]{martin07,martin08,sg07,munoz18b}, Crater~II is a
factor of $\sim4$ more extended than any fainter dwarf
\citep{torrealba16}, and Leo~T hosts neutral gas and recent star
formation \citep{rw08,dejong08}.  These objects more closely resemble
the previously-known dSphs and dSph/dIrrs in the Local Group.
Eridanus~II, on the other hand, is distinct from other UFDs only in
that it contains a star cluster \citep{crnojevic16b}.  Setting the
dividing line such that it lands between Eridanus~II and Leo~T is
therefore sensible, and minimizes the likelihood that future revisions
to the absolute magnitudes of any of these systems will blur the
boundary.

\section{STELLAR KINEMATICS}
\label{sec:kin}

Following their discovery, the first important step in clarifying the
nature of the UFDs was to determine their stellar velocity
dispersions.  By measuring the velocities of individual stars in
several systems, these early studies constrained their dynamical
masses and dark matter content.

\begin{figure}[h]
\includegraphics[width=6.33in]{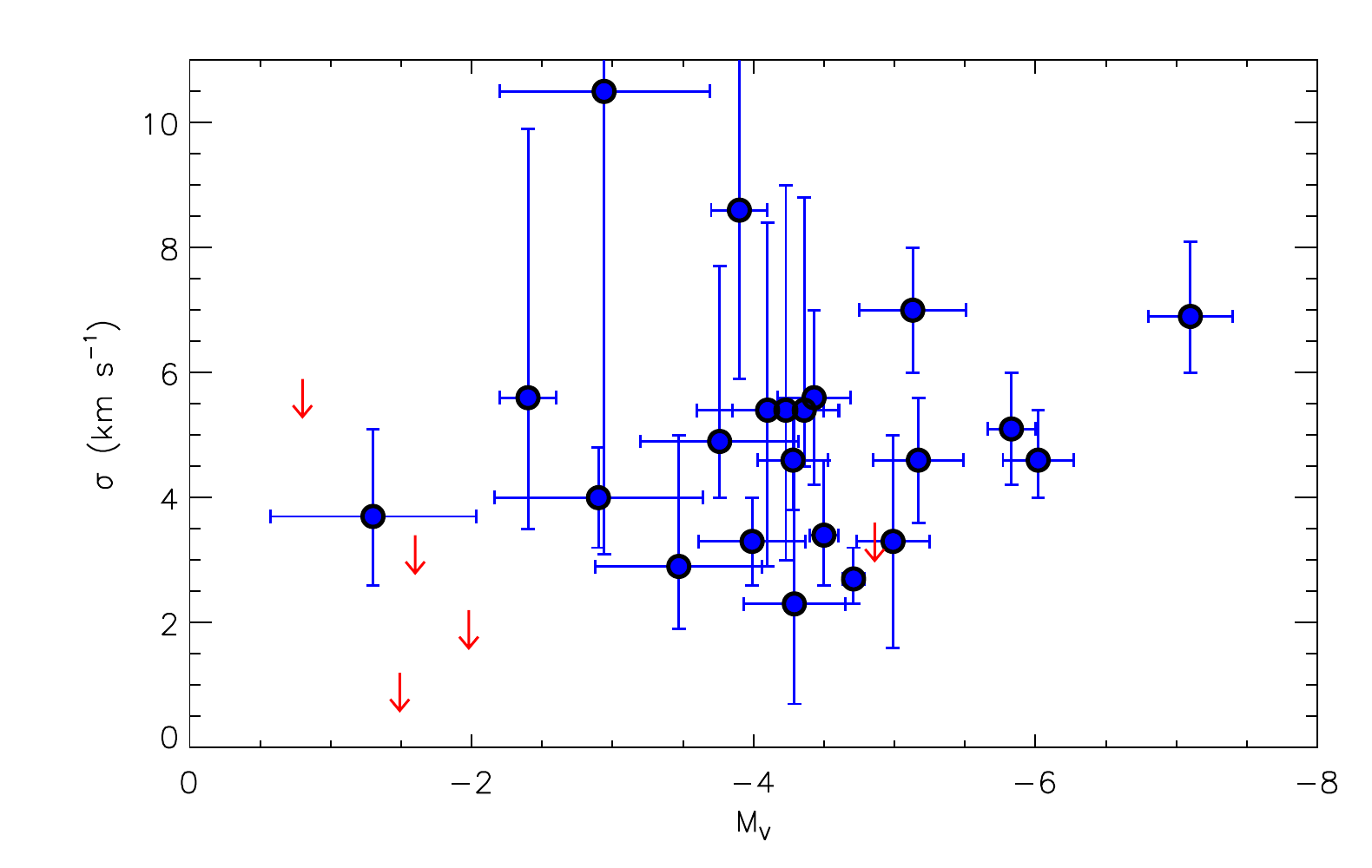}
\caption{Line-of-sight velocity dispersions of ultra-faint Milky Way
  satellites as a function of absolute magnitude.  Measurements and
  uncertainties are shown as the blue points with error bars, and
  90\%\ confidence upper limits are displayed as red arrows for
  systems without resolved velocity dispersions.  The dwarf galaxy
  data included in this plot are listed in Table 1.  Although there is
  a clear trend of decreasing velocity dispersion toward fainter
  dwarfs among the classical dSphs, in the ultra-faint luminosity
  regime there is much more scatter and any systematic trend is weak.}
\label{fig:sigma}
\end{figure}

The initial spectroscopic observations of UFDs were made by
\citet{kleyna05} for Ursa~Major~I (UMa~I) and \citet{munoz06} for
Bo{\"o}tes~I (Boo~I).  Using Keck/HIRES spectra of 5 stars,
\citeauthor{kleyna05} measured a velocity dispersion of $\sigma =
9.3^{+11.7}_{-1.2}$~\kms.  \citeauthor{munoz06} determined a velocity
dispersion of $\sigma = 6.6 \pm 2.3$~\kms\ from WIYN/Hydra spectra of
7 Boo~I stars.  These two systems have luminosities of 9600 and
21900~L$_{\odot}$, respectively.  If the stellar mass-to-light ratio
is $\approx2$~M$_{\odot}$/L$_{\odot}$ (as expected for an old stellar
population with a standard initial mass function), then the expected
velocity dispersions from the stellar mass alone would be
$\lesssim0.1$~\kms\ (making use of the \citealt{wolf10} mass
estimator).  In both cases, such a small velocity dispersion can be
ruled out at high significance, demonstrating that under standard
assumptions UFDs cannot be purely baryonic systems.  Similar
conclusions quickly followed for the remaining ultra-faint dwarfs
based on analyses of Keck/DEIMOS spectroscopy by \citet{martin07} and
\citet{sg07}.  At the present, velocity dispersion measurements or
limits have been obtained for 27 out of 42 confirmed or candidate
UFDs.  All of the available kinematic data are displayed in
Figure~\ref{fig:sigma}.

\subsection{Mass Modeling and Dark Matter Content}

\subsubsection{Assumptions Required for Determining Masses}

The results shown in Figure~\ref{fig:sigma} are simply measurements:
the observed dispersion of the radial velocities for the set of stars
in each dwarf for which spectra were obtained.  In order to translate
these velocity dispersions into dynamical masses, several assumptions
must be made.  First, no inference can be drawn about the mass of a
system unless it is in dynamical equilibrium.  If a dwarf galaxy has
experienced, for example, a recent tidal shock, then its present
velocity dispersion may not be a reliable indicator of its mass.
Recent proper motion measurements show that many of the ultra-faint
dwarfs are indeed close to their orbital pericenters, but those
pericentric passages occur at typical distances of nearly 40~kpc away
from the Galactic center, lessening their impact \citep{simon18}.
While the assumption of equilibrium deserves further attention in
modern high-resolution simulations, earlier studies indicate that even
when dwarf galaxies have been tidally disturbed their velocity
dispersions do not change substantially, and the instantaneous
dispersion remains a good barometer of the bound mass
\citep{oh95,pp95,munoz08}.

Second, unless spectroscopy of a dwarf is obtained over multiple,
well-separated observing runs, it must be assumed that binary stars
are not inflating the observed velocity dispersion above its true
value.  The influence of binary stars may be particularly concerning
given the recent suggestion that the binary fraction is quite large at
low metallicities \citep{moe18}.  Several individual binary stars have
been detected in UFDs
\citep[e.g.,][]{frebel10,koposov11,koch14,ji16b,kirby17,li18}, and the
binary population of Segue~1 was evaluated statistically by
\citet{martinez11} and \citet{simon11}.  Only the binary system in
Hercules identified by \citet{koch14} has an orbit solution (with
period $135.28 \pm 0.33$~d and velocity semi-amplitude $14.48 \pm
0.82$~\kms), but the few other UFD binaries with detected velocity
variability appear to have semi-amplitudes of $\sim10-20$~\kms\ and
periods $\lesssim1$~yr as well \citep{ji16b,kirby17,li18}.
\citet{fsk14} also found indirect evidence of binarity for a star in
Segue~1 based on its chemical abundances, which are best explained by
mass transfer from a (formerly) more massive companion star.

In the classical dSphs, a number of studies have shown that binary
stars do not significantly inflate the observed velocity dispersions
\citep[e.g.,][]{olszewski95,olszewski96,vogt95,hargreaves96,kleyna02,minor10,spencer17}.
It has also been suggested, however, that the effect of binaries may
be larger in UFDs given their smaller intrinsic velocity dispersions
\citep{mc10,spencer17}.  While that is certainly true in principle,
observationally most UFD data sets do not seem to be significantly
affected by binaries.  For example, removing the radial velocity
variables from the sample of Boo~I stars analyzed by \citet{koposov11}
changes the velocity dispersion by only $\sim3$\%.\footnote{Here we
  are modeling the Boo~I velocity distribution as a single Gaussian
  for simplicity.  \citet{koposov11} argued that the data are better
  described by a two-component model, with a majority of the stars in
  a cold $\sigma = 2.4^{+0.9}_{-0.5}$~\kms\ component and $\sim30$\%
  in a hotter component with $\sigma \approx 9$~\kms, but they were
  not able to rule out a single-Gaussian model.}  Similarly,
\citet{martinez11} and \citet{simon11} corrected the effects of
binaries in Segue~1 with Bayesian modeling of a multi-epoch radial
velocity data set and found that the binary-corrected velocity
dispersion agrees within the uncertainties with the uncorrected
dispersion.  Other recent studies have also included multi-epoch
velocity measurements, either finding no obvious binaries
\citep{simon17} or removing the binaries before computing velocity
dispersions \citep{li18}.  On the other hand, there are at least two
examples of binary stars indeed biasing the derived velocity
dispersions of UFDs: \citet{ji16b}, \citet{venn17}, and
\citet{kirby17} showed that Bo{\"o}tes~II (Boo~II) and Triangulum~II
(Tri~II) each contain a bright star in a binary system that was
responsible for substantially increasing the velocity dispersions
determined by \citet{koch09} for Boo~II and by \citet{martin16a} and
\citet{kirby15b} for Tri~II.  In both of these cases, the influence of
the binary was magnified by the very small samples of radial
velocities available (5 stars in Boo~II and $6-13$ stars in Tri~II).
These results indicate that while binary stars do not significantly
change the velocity dispersions of most ultra-faint dwarfs, studies
consisting of single-epoch velocity measurements of small numbers of
stars should be interpreted with caution.

Finally, an often unstated assumption is that samples of ultra-faint
dwarf member stars are free from contamination by foreground Milky Way
stars.  Contamination is a particularly tricky issue for galaxies with
velocities close to the median velocity of Milky Way stars along that
line of sight (e.g., Willman~1, Hercules, and Segue~2\footnote{We take
  this opportunity to note that the standard nomenclature for new
  stellar systems discovered in the Local Group for the past several
  decades has been that dwarf galaxies are named after the
  constellations in which they are located, while globular clusters
  are named after the survey in which they were found or the author
  who identified them.  When multiple discoveries are made in a single
  constellation or survey, dwarfs are usually numbered with Roman
  numerals and globular clusters with Arabic numbers.  One drawback of
  this convention is that it is no longer always obvious when an
  object is discovered how to classify it.  Consequently, Willman~1
  and Segue~1 and 2 were named as if they were globular clusters and
  then later realized to be dwarf galaxies.  The community now appears
  to be hopelessly confused about whether their numbering should be
  Roman or Arabic (the answer is Arabic; once a name is established it
  is not worth changing).  The question going forward is whether past
  naming conventions should be continued, if new conventions should be
  adopted, or if temporary names should be used until a robust
  classification is available.}), although incorrectly identified
members are possible in any dwarf.  Because stars that could be
mistaken for UFD members must have velocities close to the systemic
velocity of the dwarf, the effect of such contaminants is likely more
severe for the derived metallicity distribution than the velocity
dispersion \citep[e.g.,][]{siegel08,kirby17}.  Several examples of
erroneously determined UFD members are available in the literature.
\citet{simon18} demonstrated that stars previously classified as
members of Ursa~Major~I (UMa~I) by \citet{kleyna05} and \citet{sg07}
and Hydrus~I (Hyi~I) by \citet{koposov18} have \emph{Gaia} proper
motions that strongly disagree with the remaining members.  Removing
these stars from the member samples reduces the UMa~I velocity
dispersion from $7.6 \pm 1.0$~\kms\ to $7.0 \pm 1.0$~\kms\ and has no
effect on the measured dispersion of Hyi~I.  Similarly,
\citet{frebel10} obtained a high-resolution spectrum of a star
identified by \citet{sg07} as an Ursa~Major~II (UMa~II) member but
found that its surface gravity was not consistent with that
classification; the velocity dispersion of UMa~II is not significantly
changed by the exclusion of this star.  \citet{aden09} argued using
Str{\"o}mgren photometry that the Hercules member sample from
\citet{sg07} was contaminated by several non-member stars, but the
derived velocity dispersions are only $1.1\sigma$ apart.

For the classical dSphs, where large member samples of hundreds to
thousands of stars are generally available, a common method for
dealing with foreground contamination is to make use of membership
probabilities for each star \citep[e.g.,][]{walker09b,caldwell17}.
These probabilities are determined via a multi-component model of the
entire data set, e.g., assuming Gaussian velocity and metallicity
distributions and a \citet{plummer11} radial profile for the dwarf
galaxy.  The global properties of the dwarf can then be computed using
the individual membership probabilities as weights, with a star with a
membership probability of 0.5 counting half as much as a certain
member with a probability of 1.0.  In the limit where there are many
stars with intermediate membership probabilities ($0.1 \lesssim p_{\rm
  mem} \lesssim 0.9$), some of which are genuine members and some of
which are foreground stars, the reduced contributions of actual
members and the increased contributions of contaminants can reasonably
be assumed to cancel out so that the derived properties of the system
are accurate.  It is less clear, however, that this statistical
approach works well when applied to the small data sets typical for
UFDs.  For example, the stars with ambiguous membership status are
likely to be those that are outliers from the remainder of the
population in velocity and/or metallicity.  In reality, of course,
each such star is either a member of the dwarf or not.  If there is
only a single star in this category, probabilistically including it
as, say, 0.5 member stars may substantially change the inference on
the velocity or metallicity dispersion of the system.  In the shot
noise-limited regime, a better approach to deal with outliers may be
simply to report the derived values with and without the questionable
star(s) included, acknowledging the resulting uncertainty.
Fortunately, the advent of \emph{Gaia} astrometry should make it
possible to correctly classify most stars whose membership would
previously have been uncertain, reducing the importance of this issue
going forward.

\subsubsection{Dynamical Masses and Dark Matter}

Once the assumptions of dynamical equilibrium and minimal
contamination by binary stars and foreground stars are made, the
observed velocity dispersion can be used to constrain the mass of a
dwarf galaxy.  Early work
\citep[e.g.,][]{kleyna05,munoz06,martin07,sg07} relied on the method
of \citet{illingworth76} for determining globular cluster masses, as
applied by \citet{mateo98} to dSphs.  The Illingworth formula is based
on the dynamical model developed by \citet{king66}, again for globular
clusters.  As discussed by \citet{wolf10}, several key assumptions of
this method fail (or may fail) in the case of dwarf galaxies.  In
order of increasing severity, these assumptions include that (1) the
velocity dispersion is independent of radius, (2) the velocity
dispersion is isotropic, and (3) the mass profile follows the light
profile.  An alternate formalism is therefore needed; in particular,
one in which the mass is not assumed to be distributed in the same way
as the light.

\citet{wolf10} showed that for dispersion-supported stellar systems
with unknown velocity anisotropy, the mass that is most tightly
constrained by stellar velocity measurements is the mass enclosed
within the three-dimensional half-light radius of the system, $M_{1/2}
= M(<r_{1/2,3{\rm D}})$.  This approach still requires that the
velocity dispersion profile be approximately flat in the measured
region (which is generally the case in the dwarf galaxies for which
that measurement can be made), but does not assume anything about the
shape of the anisotropy profile or the mass distribution.  According
to \citet{wolf10},

\begin{equation}
M_{1/2} = 930 \left( \frac{\sigma}{{\rm km}~{\rm
    s}^{-1}}\right)^{2} \left(\frac{R_{1/2}}{\rm pc}\right)~{\rm M}_{\odot},
\label{eq:mhalf}
\end{equation}

\noindent
where $\sigma$ is the velocity dispersion and $R_{1/2}$ is the
projected two-dimensional half-light radius.\footnote{Similar
  relations have been derived by \citet{walker09} and
  \citet{errani18}.}  (One can also write this relation in terms of
the deprojected three-dimensional half-light radius, $r_{1/2}$, but
that is less convenient since $R_{1/2}$ is what can be measured
directly.  For many light profiles the two are simply related by
$r_{1/2} = \frac{4}{3}R_{1/2}$, as shown by \citealt{wolf10}.)

\begin{figure}[t!]
\includegraphics[width=6.33in]{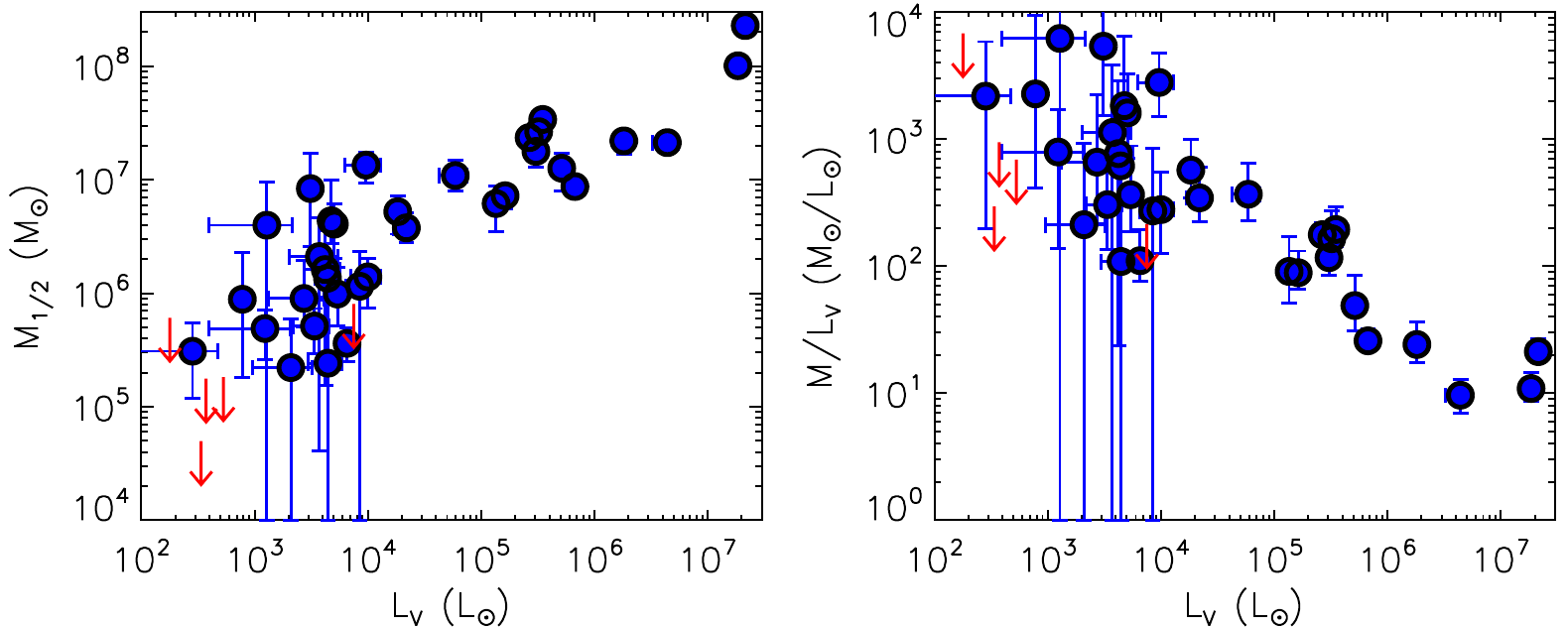}
\caption{(\emph{left}) Dynamical masses of ultra-faint Milky Way satellites
  as a function of luminosity.  (\emph{right}) Mass-to-light ratios within
  the half-light radius for ultra-faint Milky Way satellites as a
  function of luminosity.  Measurements and uncertainties are shown as
  the blue points with error bars, and mass limits determined from the
  90\%\ confidence upper limits on the dispersion are displayed as red
  arrows for systems without resolved velocity dispersions.  The dwarf
  galaxy/candidate data included in this plot are listed in
  Table 1. }
\label{fig:mhalf}
\end{figure}

The dynamical masses determined with Equation~\ref{eq:mhalf} are
displayed in Figure~\ref{fig:mhalf}.  Every UFD for which the velocity
dispersion has been measured has a mass of at least
$10^{5}$~M$_{\odot}$ within its half-light radius.  Among the five
systems with only upper limits on the dispersion available, all but
Tucana~III (Tuc~III) are consistent with such masses as well.  In
comparison, the luminosities are a factor of $\sim100$ or more
smaller.  Given that the stellar mass-to-light ratio is
$\sim2$~\msun/\lsun\ for an old stellar population with a
\citet{salpeter55} initial mass function\footnote{A \citet{kroupa01}
  or other shallower IMF \citep[e.g.,][]{geha13} has an even smaller
  stellar mass-to-light ratio.}  \citep*[e.g.,][]{martin08}, it is
clear that nearly all of the UFDs have masses that are dominated by
something other than their stars.

\subsubsection{Galaxies for which Published Kinematics May Not
  Reliably Translate to Masses}
\label{sec:caution}

The reported stellar kinematics and corresponding masses of UFDs often
seem to be regarded as having uniform reliability, especially by those
other than the original observers.  In fact, however, there are wide
variations from galaxy to galaxy in how well determined the internal
kinematics are.  The size of the member sample, the quality of the
individual velocity measurements, and the evolutionary history of the
object in question all influence the degree to which accurate
dynamical inferences can be made.  Some specific objects that should
be treated with caution include:

\begin{itemize}

\item \emph{Willman~1} --- Although \citet{willman11} identified a
  sizable sample of 40 likely Willman~1 member stars with high-quality
  Keck/DEIMOS velocity measurements, the internal kinematics of
  Willman~1 defy straightforward interpretation.  The stars closest to
  the center of Willman~1 differ in velocity by 8~\kms\ from the
  remainder of the system, and their velocity dispersion is consistent
  with zero.  While it is possible that this configuration is simply
  the result of small number statistics, it could also indicate that
  the assumption of dynamical equilibrium is not valid for Willman~1.
  On the other hand, as already pointed out by \citet{willman11},
  tidal forces would not obviously cause such a velocity distribution.

\item \emph{Boo~II} --- \citet{koch09} measured a velocity dispersion
  of $10.4 \pm 7.5$~\kms\ for Boo~II using Gemini/GMOS spectra of 5
  member stars.  The small sample of members and the presence of at
  least one binary star, as mentioned above, combine to compromise
  this measurement \citep{ji16b}.  A dedicated study of a larger set
  of Boo~II stars will likely show that the velocity dispersion of
  Boo~II is substantially smaller, in line with those of the other
  UFDs.

\item \emph{UMa~II} --- UMa~II has been the subject of two kinematic
  studies: \citet{martin07} determined a velocity dispersion of
  $7.4^{+4.5}_{-2.8}$~\kms\ from Keck/DEIMOS spectroscopy of 11
  members, and \citet{sg07} measured a dispersion of $6.7 \pm
  1.4$~\kms, also with Keck/DEIMOS, for a sample of 20 members.  The
  large velocity dispersion of UMa~II, combined with its relatively
  close distance, have been used to argue that it is one of the most
  promising targets for the indirect detection of dark matter
  \citep[e.g.,][]{ahnen18}.  One of the \citeauthor{sg07} members is
  now known to be a foreground contaminant (see above), but this star
  does not impact the velocity dispersion.  Subsequent to the velocity
  dispersion measurements, one binary star was detected in UMa~II by
  \citet{frebel10}.  Removing these two stars from the sample reduces
  the velocity dispersion to $5.6 \pm 1.4$~\kms, but we also note that
  most of the other UMa~II members have not been checked for binarity.
  Furthermore, UMa~II is the only UFD for which the typical Plummer or
  exponential radial profiles fail to provide a good fit
  \citep{munoz10,munoz18b}.  Its unusual profile may be consistent
  with tidal disruption \citep*{munoz10}, but on the other hand, the
  orbit of UMa~II has a pericenter of $39^{+2}_{-3}$~kpc and a long
  period of $\sim3.5$~Gyr, suggesting that the dwarf has completed at
  most 3 orbits around the Milky Way and has never approached closely
  enough for its stars to be tidally stripped \citep{simon18}.
  Verifying that UMa~II indeed has a larger velocity dispersion than
  other UFDs and is in equilibrium will require spectroscopy of a
  larger sample of stars over a wider area, as well as repeat velocity
  measurements to check for the binarity of known members.

\item \emph{Boo~I} --- \citet{koposov11} presented a high-quality
  VLT/FLAMES kinematic data set for Boo~I.  They showed that, unlike
  other UFDs, its line of sight velocity distribution is best
  described by two distinct components, one with a dispersion of
  $2.4^{+0.9}_{-0.5}$~\kms\ and the other with a dispersion of
  $\approx9$~\kms.  These components are reminiscent of the multiple
  populations with separate radial, metallicity, and velocity
  distributions seen in some of the more luminous dSphs
  \citep[e.g.,][]{tolstoy04,battaglia06,battaglia11}.  However, in
  Boo~I the available stellar samples are too small to confidently
  detect the photometric or chemical signatures of two populations
  \citep{koposov11}.  In the absence of a half-light radius to
  associate with each kinematic component, we emphasize that using one
  of the cold or hot velocity dispersions alone to calculate the mass
  of Boo~I is not valid.

\end{itemize}

\subsection{Identification as Galaxies}
\label{sec:galaxies}

Prior to the discovery of dwarf galaxies fainter than $M_{{\rm V}}
\sim -5$, dwarfs and globular clusters occupied distinct and cleanly
separated portions of the size-luminosity parameter space displayed in
Fig.~\ref{fig:mv_rhalf}.  Consequently, there was little discussion of
whether certain objects should be considered to be galaxies or
clusters; the classification of all known systems was
obvious.\footnote{The exception to this statement is the idea that
  some globular clusters, most notably $\omega$~Centauri, might be the
  remnant nuclei of tidally disrupted dwarf galaxies
  \citep[e.g.,][]{lee99,majewski00,hr00}.}

As the dwarf galaxy population grew toward lower luminosities and
smaller radii, the gap between dwarfs and globular clusters in the
size-luminosity plane disappeared, such that size alone was no longer
sufficient to determine the nature of an object.  \citet{conn18}
dubbed the region occupied by a number of ultra-faint satellites ($14
< r_{1/2} < 25$~pc) the ``trough of uncertainty'' to emphasize the
difficulty in classifying these systems.  In order to resolve the
confusion caused by the lack of an agreed-upon system for separating
galaxies from star clusters, \citet{ws12} proposed the following
definition: 

\begin{extract}
``A galaxy is a gravitationally bound collection of stars
whose properties cannot be explained by a combination of baryons and
Newton's laws of gravity.'' 
\end{extract}

\noindent
Applied to UFDs, this definition is generally interpreted as requiring
an object to have either a dynamical mass significantly larger than
its baryonic mass or a non-zero spread in stellar metallicities to
qualify as a galaxy.  The former criterion directly indicates the
presence of dark matter (for which there is no evidence in globular
clusters), while the latter indirectly suggests that the object must
be embedded in a dark matter halo massive enough that supernova ejecta
can be retained for subsequent generations of star formation.

The early SDSS UFDs were all measured to have velocity dispersions
larger than 3~\kms, implying that they are composed mostly of dark
matter and can be straightforwardly classified as galaxies
\citep{kleyna05,munoz06,martin07,sg07,geha09}.  Some disagreement
persisted for several years regarding the nature of the two
least-luminous systems, Willman~1 and Segue~1
\citep[e.g.,][]{belokurov07,siegel08,no09}, but ultimately the
combination of stellar kinematics, metallicities, and chemical
abundance measurements led to the conclusion that both are dwarfs
\citep{willman11,simon11,fsk14}.  The first object for which kinematic
classification failed entirely was Segue~2.  Despite a comprehensive
analysis of the kinematics of Segue~2, \citet{kirby13} were unable to
measure its velocity dispersion, finding $\sigma < 2.6$~\kms\ at 95\%
confidence.  With only an upper limit on the dynamical mass and
mass-to-light ratio, it cannot be confirmed that Segue~2 contains dark
matter.  However, \citeauthor{kirby13} also showed that the stars of
Segue~2 span a large range of metallicities, from ${\rm \feh} = -2.9$
to ${\rm \feh} = -1.3$, with a dispersion of 0.43~dex.  Segue~2 can
therefore still be classified as a galaxy rather than a globular
cluster on the basis of its chemical enrichment.

The discovery of larger numbers of compact ultra-faint satellites in
Dark Energy Survey and Pan-STARRS data
\citep{bechtol15,koposov15,drlica15,laevens15,laevens15b} has
increased the difficulty in classification.  For several of these
objects only upper limits on the velocity dispersion have been
obtained \citep{kirby15,kirby17,martin16b,simon17}, and in the case of
Tuc~III no metallicity dispersion is detectable in current data either
\citep{simon17}.  In such situations one must either accept the
uncertainty in the nature of some systems or rely on more
circumstantial arguments such as size, survival in a strong tidal
field \citep[e.g.,][]{simon17}, mass segregation \citep{kim15kim2}, or
chemical peculiarities .

As of this writing, the following 21 satellites can be regarded as
spectroscopically confirmed ultra-faint dwarf galaxies: Segue~2,
Hydrus~1, Horologium~I, Reticulum~II, Eridanus~II, Carina~II,
Ursa~Major~II, Segue~1, Ursa~Major~I, Willman~1, Leo~V, Leo~IV,
Coma~Berenices, Canes~Venatici~II, Bo{\"o}tes~II, Bo{\"o}tes~I,
Hercules, Pegasus~III, Aquarius~II, Tucana~II, and Pisces~II.
Satellites that may be dwarfs but for which either no spectroscopy has
been published or the data are inconclusive include: Tucana~IV,
Cetus~II, Cetus~III, Triangulum~II, DES~J0225+0304, Horologium~II,
Reticulum~III, Pictor~I, Columba~I, Pictor~II, Carina~III, Virgo~I,
Hydra~II, Draco~II, Sagittarius~II, Indus~II, Grus~II, Grus~I,
Tucana~V, Phoenix~II, and Tucana~III.  The most extended of these
objects, such as Tucana~IV, Cetus~III, Columba~I, and Grus~II, are
perhaps the most likely to be dwarfs given their large radii.  We have
not included Bo{\"o}tes~III, which is likely the remnant of a dwarf,
in either category because it is not clear whether it is still a bound
object \citep{grillmair09,carlin09,cs18}.

The problem of determining the nature of the faintest and most compact
Milky Way satellites will only become more severe in coming years as
surveys become sensitive to even lower luminosity, lower surface
brightness, and more distant stellar systems.  Spectroscopic follow-up
of the satellites discovered by LSST will require massive investments
of telescope time on either existing facilities or those currently
under consideration \citep{najita16}.

\section{METALLICITIES AND CHEMICAL ABUNDANCES}
\label{sec:metallicity}

The metallicities of stars in UFDs are important both for classifying
them as galaxies (Section~\ref{sec:galaxies}) and for connecting them
to the broader field of galaxy formation (Section~\ref{sec:highz}).
Fortunately, the same spectra of individual stars from which the
stellar kinematics are determined can often be used to measure
metallicities.  With spectral synthesis techniques developed over the
last decade and other methods, abundances of several elements other
than iron can also be obtained from medium-resolution spectra of dwarf
galaxy stars
\citep[e.g.,][]{kirby09,norris10,kirby11,vargas13,koposov15b}.  Mean
metallicities based on such data have now been obtained for 26 out of
42 confirmed/candidate UFDs.  Detailed chemical abundance patterns
generally require observations at higher spectral resolution, which
are challenging for dwarf galaxies because even their brightest stars
are usually rather faint.

The first spectroscopic metallicity measurements for ultra-faint
dwarfs were provided by \citet{munoz06}, \citet{martin07}, and
\citet{sg07}.  Collectively, these studies showed that the ultra-faint
dwarfs have very low metallicities (${\rm \feh} \lesssim -2$) and that
the stars in each object span a range in metallicity.  The latter
property distinguishes UFDs from globular clusters, and indicates both
that star formation in these objects extended for a long enough period
of time for supernova (SN) enrichment to occur and that their
gravitational potential is deep enough that not all of the supernova
ejecta can escape the system.  \citet{kirby08} used more accurate
metallicity measurements to show that many of the ultra-faint dwarfs
contain extremely metal-poor (EMP) stars with ${\rm \feh} < -3$, again
distinct from globular clusters and (at the time) classical dSphs.

UFDs are particularly appealing systems in which to study early
chemical evolution and nucleosynthesis because their small stellar
masses imply that they have hosted relatively few SN explosions.  That
fact, combined with the short time periods during which they were
forming stars (see Section~\ref{sec:stellar_pops}), means that UFDs
may preserve the unpolluted chemical signatures of small numbers of
nucleosynthetic events \citep[][]{bh10,karlsson13}, perhaps even
individual explosions \citep[e.g.,][]{fb12,ji15}.  \citet{koch08}
began the process of analyzing the detailed chemical abundances of UFD
stars with high-resolution Magellan/MIKE spectra of two stars in
Hercules.  \citet{frebel10} and \citet{norris10b,norris10c} extended
this effort to more dwarfs and lower metallicities.  Even from these
earliest studies, it was clear that the UFDs are enhanced in $\alpha$
elements such as oxygen, magnesium, calcium, and silicon, and
unusually deficient in neutron-capture elements including barium and
strontium, as detailed further in Section~\ref{sec:abund_patterns}.

\subsection{The Mass-Metallicity Relation}

\begin{figure}[h]
\includegraphics[width=6.33in]{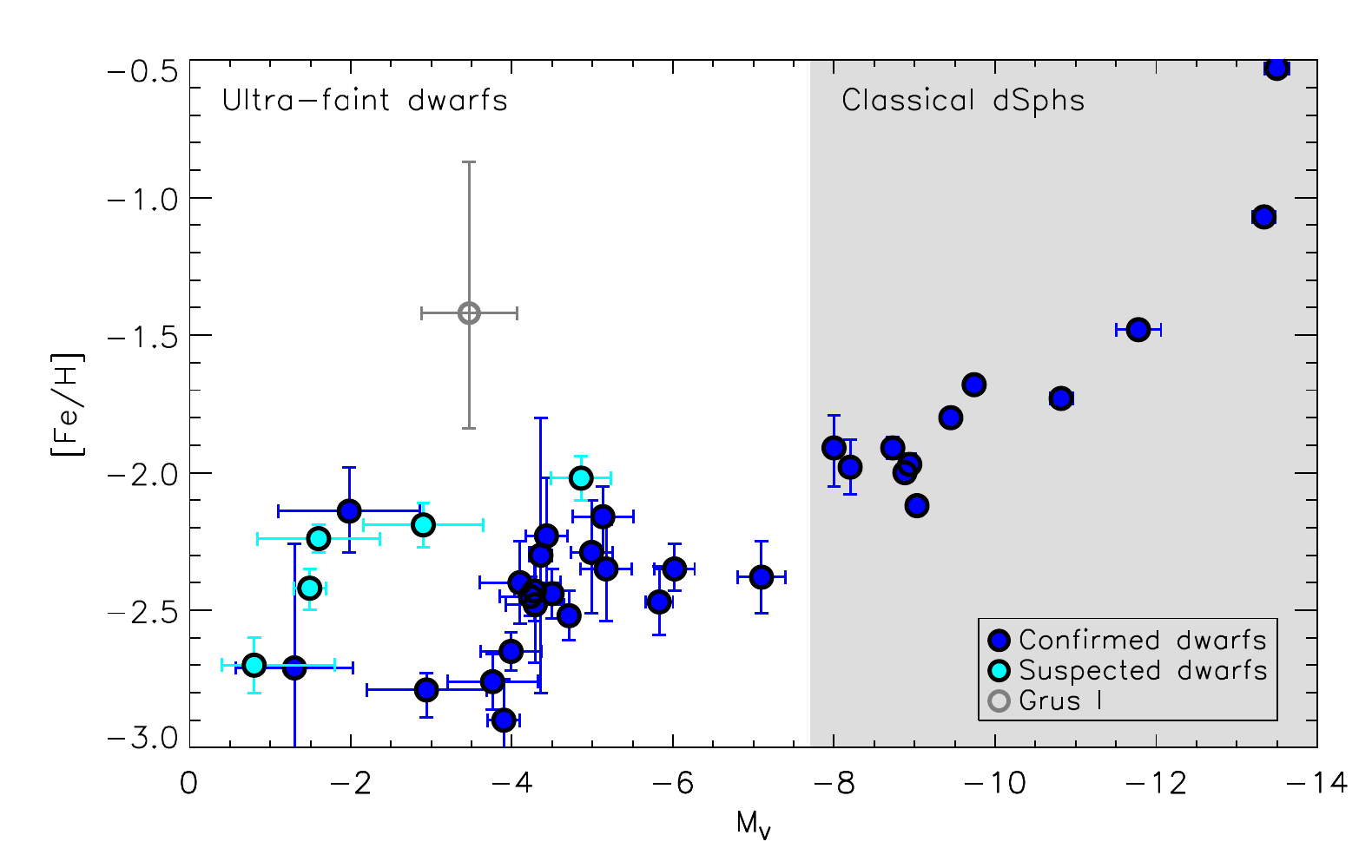}
\caption{Mean stellar metallicities of Milky Way satellites as a
  function of absolute magnitude.  Confirmed dwarf galaxies are
  displayed as dark blue filled circles, and objects suspected to be
  dwarf galaxies but for which the available data are not conclusive
  are shown as cyan filled circles.  Grus~I, for which there is no
  published classification, is shown as an open gray circle.  The
  error bars in the vertical direction indicate the uncertainty on the
  mean metallicity of each object.  The dwarf galaxy/candidate data
  included in this plot are listed in Table 1.  The overall
  relationship between metallicity and luminosity is clear, although
  the scatter is large at the faint end.}
\label{fig:feh}
\end{figure}

A correlation between the stellar mass or luminosity of a galaxy and
its mean metallicity has been known for decades \citep[e.g.,][and
  references therein]{tremonti04}.  \citet{sg07} and \citet{kirby08}
showed that such a relationship also exists even in the ultra-faint
dwarf regime, more than five orders of magnitude in luminosity below
the galaxies examined by \citeauthor{tremonti04}  \citet{kirby13b}
carefully quantified the stellar mass-metallicity relation for Local
Group dwarfs, demonstrating that a single relation holds for all types
of dwarf galaxies throughout the Local Group: 

\begin{equation}
{\rm \feh} = (-1.68 \pm 0.03) + (0.29 \pm
0.02)\log{\left(\frac{L_{V}}{10^{6}~L_{\odot}}\right)},
\end{equation}

\noindent
with a standard deviation around the fit of only 0.16~dex.  Including
measurements made since 2013 for a larger sample of Milky Way
satellites (see Figure~\ref{fig:feh}), we find a best fit consistent
with that reported by \citet{kirby13b}, although with an increased
intrinsic scatter of $\sim0.25$~dex primarily attributable to the
faintest dwarfs.

The existence of a tight correlation between luminosity and
metallicity argues against severe tidal stripping of the stellar
components of dwarf galaxies.  Tidal stripping reduces the luminosity
of a system without significantly changing its
metallicity.\footnote{In the case of a metallicity gradient with the
  most metal-rich stars near the center of the galaxy, stripping would
  actually increase the overall metallicity slightly as low
  metallicity stars in the outskirts are preferentially stripped.}
Stripping will therefore tend to increase the scatter in the
correlation; indeed, the two dwarfs known to be stripped because of
the presence of substantial tidal tails, Sagittarius and Tuc~III, lie
$\sim0.8$~dex and $\sim0.5$~dex above the correlation.  The fact that
the correlation remains in place therefore puts an upper limit on the
amount of stripping that could have occurred for the bulk of the dwarf
galaxy population.  However, since the dark matter halos of galaxies
are much more extended than their stars, a large fraction of the dark
matter in dSphs and UFDs could be removed without affecting the
luminosity-metallicity relation.

The reader may observe that the scatter in the luminosity-metallicity
relation appears to increase substantially for UFDs around $M_{{\rm
    V}} \gtrsim -5.5$, and even more so at $M_{{\rm V}} \gtrsim -4.0$.
This increased dispersion could be interpreted as evidence that the
faintest dwarfs have suffered more stripping than the classical dSphs.
Alternatively, (underestimated) observational uncertainties and errors
may be responsible for some or all of the larger scatter at the lowest
luminosities.  In particular, the metallicities of some of the
outliers above the relation are currently determined from only two
member stars (e.g., Willman~1 and Tri~II).  If the brightest stars in
those systems happen not to be representative of the overall
metallicity distribution then the derived mean metallicity will be
incorrect.  The most significant outlier is Grus~I, which is reported
by \citet{walker16} to contain four stars (out of seven measured) with
${\rm \feh} > -1.4$.  No other UFD contains so many metal-rich stars,
suggesting that some of them are probably foreground contaminants and
that the system is actually more metal-poor.

\subsection{Metallicity Distribution Functions}

In contrast to the classical dSphs, relatively little work has been
done on the metallicity distribution functions (MDFs) in ultra-faint
dwarfs.  This lack of attention is largely a result of the small
samples of metallicity measurements typically available for UFDs.  The
best-studied object is Boo~I, for which \citet{norris10} derived an
MDF with 16 stars and \citet{lai11} determined an MDF with 41 stars.
The shape of the Boo~I MDF is qualitatively similar to those obtained
by \citet{kirby11} for the classical dSphs, although with a narrower
peak.  \citet{kirby08} showed that the combined MDF of seven UFDs and
CVn~I is very similar to the MDF of the Milky Way halo over the
metallicity range from ${\rm \feh} = -2$ to ${\rm \feh} = -3.5$.
\citet{brown14} determined MDFs for six UFDs, finding suggestions of
multiple peaks in the metallicity distributions in several cases (most
notably Boo~I and Hercules).  However, given the sparseness of the
data for most galaxies, few authors have attempted to draw conclusions
about the evolutionary history of UFDs from the observed MDFs (see
Section~\ref{sec:chem_evol}).  \citet{lai11} found that the Extra Gas
model of \citet{kirby11} provides the best fit to the Boo~I MDF, with
most of the stars forming from an accreted reservoir of pristine gas,
although the alternative \citeauthor{kirby11} models (Pre-Enriched and
Pristine) also fit the data acceptably well.  The effective yield
derived by \citet{lai11} for any of the three models continues the
trend found by \citet{kirby11} of decreasing effective yield with
decreasing stellar mass.

\subsection{Chemical Evolution Histories}
\label{sec:chem_evol}

Beyond the MDF, the most basic feature of galactic chemical evolution
is the dependence of the abundance of $\alpha$ elements on
metallicity.  At low metallicity, high [$\alpha$/Fe] ratios are
observed, while more metal-rich stars have low [$\alpha$/Fe] ratios.
This behavior results from the different chemical yields from
different types of SNe.  Core-collapse SNe from massive stars explode
quickly after star formation occurs, producing large quantities of
$\alpha$ elements.  As time passes, Type Ia SNe begin to explode,
producing primarily iron-peak elements and thereby lowering the
[$\alpha$/Fe] ratio \citep{tinsley79}.

The timing of the transition between chemical enrichment dominated by
core-collapse SNe and SNe Ia varies from galaxy to galaxy because it
depends on the star formation rate \citep[e.g.,][]{venn04,kirby11}.
In the compilation of 7 UFDs by \citet{vargas13}, this transition
appears to occur rather sharply at ${\rm \feh} = -2.3$ when the data
for the entire sample are combined.  The observation of high
[$\alpha$/Fe] at ${\rm \feh} < -2.3$ and $\sim$solar [$\alpha$/Fe] at
${\rm \feh} > -2.3$ is interpreted in the standard picture as evidence
that Type Ia SNe began to contribute significantly to chemical
enrichment at ${\rm \feh} \approx -2.3$.  In that case, star formation
in UFDs must have continued for $\gtrsim100$~Myr so that some SNe~Ia
exploded before the cessation of star formation
\citep[e.g.,][]{vargas13}.  However, \citet{jeon17} suggested instead
that the lack of a clear knee in the [$\alpha$/Fe] vs. \feh\ diagram
indicates that the UFDs were predominantly enriched by core-collapse
SNe.

Only one galaxy, Segue~1, shows no evidence for a change in
[$\alpha$/Fe] over a broad range in metallicity
\citep{vargas13,fsk14}.  This abundance pattern is consistent with the
one-shot enrichment scenario of \citet[][although see
  \citealt{webster16} for alternative possibilities]{fb12}, with star
formation in Segue~1 likely lasting for less than a few hundred Myr
and ending before any SNe~Ia occurred.

Analytical chemical evolution models can provide insight into star
formation and nucleosynthesis processes in galaxies \citep[e.g.,][see
  \citealt{at76} and \citealt{nomoto13} for reviews]{ss72}.  Thus far,
such models have only been applied to two UFDs, Hercules and Boo~I
\citep{vincenzo14, romano15}.  By simultaneously fitting the observed
stellar masses, MDFs, and [$\alpha$/Fe] ratios, \citet{vincenzo14}
showed that the UFDs formed with smaller gas reservoirs and star
formation efficiencies a factor of $\sim10$ lower than the classical
dSphs.  In agreement with previous results from the classical dSphs
\citep[e.g.,][]{kirby11b}, \citet{vincenzo14} found that most of the
gas and metals are removed from the galaxies by galactic winds,
although \citet{romano15} concluded that gas removal by tidal and
ram-pressure stripping is more likely for Boo~I.  Extending these
models to a larger sample of UFDs covering a wider range of parameter
space would be very interesting, but requires increased numbers of
metallicity and $[\alpha$/Fe] measurements to be feasible.

Several recent numerical studies have explored the chemical evolution
of UFD galaxies via hydrodynamic simulations.  Using idealized models,
\citet{webster14} and \citet{bh15} showed that dark matter halos as
small as $10^{7}$~\msun\ can retain gas after SN explosions, while
less massive halos are evacuated after a single SN.  They also found
that only SNe near the center of a galaxy have a significant impact on
its gas content; most of the energy from SNe that explode in the
outskirts is lost to the intergalactic medium.  In these models, the
observed behavior of [$\alpha$/Fe] as a function of metallicity from
\citet{vargas13} can be reproduced if the duration of star formation
is a few hundred Myr.  \citet{webster15} extended this work by
examining the effect of different star formation histories.  They
concluded that multiple well-separated bursts of star formation, as
modeled by, e.g., \citet{brown14}, produce more extremely metal-poor
stars and fewer low [$\alpha$/Fe] stars than observed.  However, a
model with continuous star formation (with brief pauses as the gas is
heated by SNe) and ongoing self-enrichment provides a reasonable match
to the data.  More sophisticated simulations have been carried out by
\citet{jeon17}, who studied the formation of several UFDs in a
cosmological context with a chemical reaction network.  \citet{jeon17}
demonstrated that a combination of reionization and SN feedback is
necessary to quench star formation in these objects.  Each of the
dwarfs they simulated experienced several mergers at early times, with
$\sim10-20$\%\ of the stars forming outside the main progenitor halo.
In these models, the lowest metallicity UFD stars formed in halos that
were enriched by Population~III SN explosions in neighboring halos,
while stars at ${\rm \feh} \gtrsim -3$ primarily formed in situ with
enrichment dominated by Population~II SNe.  As with the simpler models
of \citet{webster15}, the chemical abundances predicted by the
\citeauthor{jeon17} simulations ([C/Fe] and [$\alpha$/Fe]) are in
reasonable agreement with observed UFDs.

\subsection{Chemical Abundance Patterns}
\label{sec:abund_patterns}

Early studies of the detailed chemical abundance patterns of stars in
UFDs focused on broad trends as a function of metallicity, which for
the most part resemble the abundance trends of Milky Way halo stars in
the same metallicity range
\citep[e.g.,][]{koch08,frebel10,norris10b,norris10c}.  Ultra-faint
dwarf stars are enhanced in $\alpha$ elements by $\sim0.3$~dex, have
Cr abundances that increase linearly with metallicity, are sometimes
enhanced in carbon, and have low abundances of neutron-capture
elements.  Stars that are outliers from these general results in
specific abundance ratios do exist, but their frequency does not seem
to be high \citep[e.g.,][]{vargas13}.

\subsubsection{Typical Ultra-Faint Dwarfs}

\begin{figure}[t!]
\includegraphics[width=6.33in]{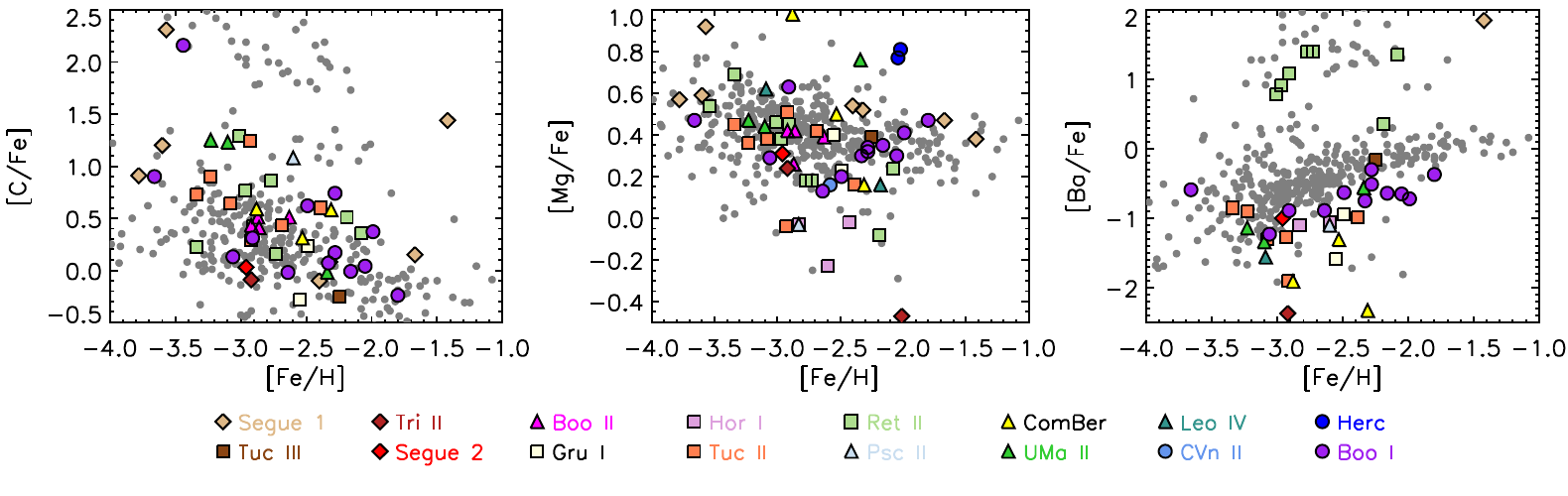}
\caption{Chemical abundance patterns of stars in UFDs.  The left,
  middle, and right panels show [C/Fe], [Mg/Fe], and [Ba/Fe] ratios as
  a function of metallicity, respectively.  UFD stars are plotted as
  colored diamonds, squares, triangles, and circles, as listed in the
  legend at bottom.  The UFD data have been adopted from
  \citet{koch08}, \citet{feltzing09}, \citet{frebel10,frebel16},
  \citet{simon10}, \citet{norris10c,norris10,norris10b},
  \citet{lai11}, \citet{gilmore13}, \citet{koch13}, \citet{fsk14},
  \citet{ishigaki14}, \citet{rk14}, \citet{ji16c,ji16b},
  \citet{francois16}, \citet{kirby17}, \citet{hansen17},
  \citet{nagasaw18}, \citet{chiti18}, \citet{spite18}, and
  \citet{ji18}.  A sample of metal-poor Milky Way halos stars from
  \citet{cohen13} and \citet{roederer14} is displayed as small gray
  circles for comparison.}
\label{fig:abundances}
\end{figure}

Chemical abundance measurements from high-resolution spectroscopy are
now available for at least one star in 16 UFDs.  This sample currently
contains more than 50 stars, with metallicities ranging from ${\rm
  \feh} = -1.4$ to ${\rm \feh} = -3.8$.  With a handful of exceptions,
the abundance patterns of different ultra-faint dwarfs closely
resemble each other \citep[see, e.g.,][]{chiti18}, such that the
galaxy in which a star is located cannot be discerned by examining its
chemical abundances (see Figure~\ref{fig:abundances}).  Some of the
examples of distinct abundance patterns include the low [$\alpha$/Fe]
ratios in Horologium~I \citep{nagasaw18} and low [Sc/Fe] ratios in
Com~Ber and possibly Segue~2 \citep{frebel10,rk14}.

For elements through the iron peak, the abundances of ultra-faint
dwarf stars closely follow the halo trend as a function of metallicity
(Fig.~\ref{fig:abundances}).  This result strongly suggests that
nucleosynthesis and chemical evolution at early times do not depend
significantly on galactic environment
\citep[e.g.,][]{tolstoy09,simon10}.  Whether the dispersion in
abundance for each element at a constant metallicity matches between
halo stars and the UFDs has not been investigated, but could be
illuminating as to early chemical evolution and star formation.  At
the lowest metallicities, a significant fraction of UFD stars have
high carbon abundances
\citep[e.g.,][]{frebel10,norris10c,salvadori15,ji16c,spite18}, again
matching previous findings for the halo.

The only clear distinction between the abundance patterns of UFDs and
halo stars is seen in the heaviest elements.  For most UFD stars these
measurements are limited to Ba and Sr, which have the strongest lines
at optical wavelengths for typical enrichment levels.  The UFD [Ba/Fe]
and [Sr/Fe] abundance ratios are not usually outside the distribution
of Ba and Sr abundances in the Milky Way halo, but most UFDs have
abundance ratios that are significantly ($\gtrsim1$~dex) lower than
the average ratios for halo stars.  Despite the much larger sample of
halo stars available in the literature, some of the lowest Ba and Sr
abundances ever measured are found in UFDs.  Very low abundances of
neutron-capture elements have been suggested as a possible
characteristic for distinguishing UFDs from globular clusters in
difficult cases \citep{ji18}; low aluminum abundances may also
separate dwarfs from clusters.

\subsubsection{Rare Enrichment Events in Ultra-Faint Dwarfs}

Perhaps the biggest surprise from chemical abundance measurements in
UFDs was the recent discovery of extreme enrichment of r-process
elements in Reticulum~II \citep{ji16,roederer16}.  Out of nine Ret~II
stars for which high-resolution spectra have been obtained, seven have
${\rm [Eu/Fe]} > 0.9$ and ${\rm [Ba/Fe]} > 0.3$ \citep{ji16}.  From
the full sample of previously-studied ultra-faint dwarf stars, none
have ${\rm [Ba/Fe]} > 0$ and only one (an s-process-enhanced binary in
Segue~1) has ${\rm [Eu/Fe]} > 0$.  In fact, the Segue~1 binary, which
was presumably contaminated with heavy elements by a companion that
went through an AGB phase, was the only previous star in which Eu had
been detected at all.  Compared to most stars in UFDs, Ret~II is
enriched in Sr and Eu by a factor of $>30$, and in Ba by a factor of
$\gtrsim100$.  As shown by \citet{ji16,ji16c} and \citet{roederer16},
the abundance patterns of the Eu-rich stars in Ret~II perfectly match
the r-process enrichment pattern seen in r-process-enhanced Milky Way
stars and the Sun.

The only viable explanation for the chemical abundances of Ret~II is
that a single nucleosynthetic event early in the history of the galaxy
produced a large quantity of r-process material
($\sim10^{-5}$~\msun\ of Eu; \citealt{ji16}).  Given that nine other
UFDs had been observed prior to Ret~II, and no sign of significant
r-process enrichment was detected in any of them, whatever produced
the neutron-capture elements in Ret~II must have been a rare
occurrence.  Ret~II does not have a large stellar mass among UFDs, so
it is very unlikely that an event would take place multiple times in
Ret~II and never in any other similar galaxies.  Therefore, the very
large overabundance of r-process elements also indicates that this
single event must have produced copious amounts of such elements.
Ordinary SNe do not have these characteristics, leaving a neutron star
merger or a magneto-rotationally driven SN as the possible sites for
r-process nucleosynthesis in Ret~II \citep{ji16,roederer16}.  As a
result of the observational evidence for r-process nucleosynthesis by
the neutron star merger GW170817
\citep[e.g.,][]{chornock17,cowperthwaite17,drout17,kasliwal17,pian17,shappee17,smartt17},
as well as the approximate agreement in the inferred rates of such
events, a neutron star merger is heavily favored as the primary origin
of r-process elements in Ret~II and other dwarf galaxies.

Subsequent to the identification of Ret~II as an r-process-rich
galaxy, \citet{hansen17} showed that the brightest star in Tuc~III is
also enhanced in r-process elements.  The rate of r-process
enhancement in UFDs can therefore be updated to two out of 14
galaxies.  The r-process abundances in Tuc~III are significantly lower
than in Ret~II, but still well above those in any other UFD.  This
abundance pattern implies that either the nucleosynthetic event in
Tuc~III produced less r-process material, it was diluted into a larger
mass of gas, or that a larger fraction of the r-process yield of the
event escaped the galaxy.

Although the abundance of r-process elements in other ultra-faint
dwarfs is several orders of magnitude lower than in Ret~II and
Tuc~III, those elements are still detected at very low levels in
almost every galaxy for which sufficient data exist
\citep[e.g.,][]{frebel10,roederer13,gilmore13,fsk14,ishigaki14,chiti18}.
If Ret~II was enriched by a single r-process event, then the only way
the same mechanism could account for much lower --- but nonzero ---
r-process abundances in other UFDs is if the gas masses of those
systems were much larger than in Ret~II or the retention fraction of
r-process ejecta were much lower.  Straightforward calculations
indicate that these possibilities are unlikely (see below).
Therefore, the natural conclusion is that a second site of r-process
nucleosynthesis exists, which produces much smaller amounts of
r-process material \citep{ji16c}.  This alternate pathway of creating
heavy elements can likely be identified with core-collapse SNe
\citep[e.g.,][]{lee13}.

Theoretical modeling of the r-process enrichment of UFDs is still in
its early stages.  However, the work that has been done from several
different directions confirms that the scenario proposed by
\citet{ji16} is both physically plausible and can quantitatively
explain the observed r-process abundances.  Specifically,
\citet{beniamini16a,beniamini16b} and \citet{beniamini18} used
analytical calculations to demonstrate that (1) binary neutron stars
can be retained in low-mass dwarfs despite possible SN kicks, (2) a
non-negligible fraction of such binaries will merge in less than
$\sim100$~Myr, (3) each nucleosynthesis event must produce
$\sim10^{-5}$~\msun\ of Eu and occur at a rate of one per
$\sim2000$~SNe, (4) up to $\sim90$\% of the r-process material
synthesized should remain in the galaxy, and (5) $\sim7$\% of UFDs
should be enriched by an r-process event.  Chemical evolution modeling
by \citet{ks16} showed that r-process synthesis by neutron star
mergers can also reproduce the distribution of [Eu/Fe] and [Ba/Fe]
abundance ratios for metal-poor stars in the Milky Way halo, under the
assumptions of a lower star formation efficiency in dwarf galaxies and
a size-dependent escape fraction for r-process material.  Hydrodynamic
simulations of UFDs by \citet{safarzadeh17} also support this picture,
showing that a neutron star merger near the center of Ret~II can
reasonably account for the observed distribution of [Eu/H] and [Fe/H]
values.  If the merger occurred in the outskirts of the galaxy, on the
other hand, lower Eu abundances that are uncorrelated with metallicity
would be expected.

\begin{landscape}
\input{dwarf_tab.tex}
\clearpage
\end{landscape}

\section{STRUCTURAL PROPERTIES}
\label{sec:struc}

A homogeneous analysis of the structural properties of the early SDSS
UFDs was provided by \citet{martin08}.  \citet{munoz18b} recently
updated this work, presenting uniform processing of deep photometry
for all UFDs known as of mid-2015 (of course, a number of new dwarfs
have been discovered since that date).  Many previous studies have
shown that the radial profiles of ultra-faint dwarfs can be accurately
described by either exponential or \citet{plummer11}
profiles\footnote{As discussed in Section~\ref{sec:caution}, the only
  UFD for which these profiles do not fit is UMa~II
  \citep{munoz10,munoz18b}.}
\citep[e.g.,][]{belokurov06,martin08,sand10}.  \citet{munoz18b}
advocated instead for S{\'e}rsic profiles, which match the observed
radial profiles more closely (and are widely used for brighter
galaxies), at the cost of one additional degree of freedom in the fit.
Confirmed UFDs have half-light radii ranging from 24~pc (Segue~1) to
295~pc (UMa~I), with candidate UFDs extending down to $15-20$~pc in a
few cases.  In comparison, the classical dSphs have half-light radii
between 170~pc (Leo~II) and 2660~pc (Sagittarius).

Apart from simply being smaller on average, it has also been suggested
that the faintest galaxies have significantly larger ellipticities
than larger systems \citep{martin08}.  Updating the samples from what
was available ten years ago, we calculate a weighted average
ellipticity for the UFDs of $0.50 \pm 0.01$, while the weighted
average ellipticity of the classical dSphs is $0.350 \pm 0.003$, in
good agreement with the statistics determined by \citet{martin08}.
However, using a two-sided Kolmogorov-Smirnov test we find that there
is a 19\% probability that the two samples are drawn from a common
distribution (as previously indicated by \citealt{sand12}).  We
therefore conclude that there is no significant evidence at present
that UFDs have more elongated shapes than more luminous dwarfs.

A recurring question regarding the structure of the faintest dwarfs is
whether their isophotes are irregular or distorted in any way, which
could suggest recent tidal stripping
\citep[e.g.,][]{zucker06b,belokurov06,okamoto12}.  Several analyses of
simulated photometric data sets of faint dwarfs have shown that these
apparently irregular shapes are the result of Poisson fluctuations in
the distribution of stars in the lowest surface brightness regions of
these systems rather than evidence for disturbed morphology
\citep{walsh08,martin08,munoz10}.

The recent discovery of the relatively luminous ($M_{{\rm V}} =
-8.2$), but extremely diffuse, Crater~II dwarf \citep{torrealba16}
highlights the possibility that the currently known population of
dwarf galaxies may be limited in surface brightness by the sensitivity
of existing photometric surveys.  Indeed, \citet{munoz18b} clearly
illustrate how the discovery of new Milky Way satellites has pushed to
lower and lower surface brightnesses as the available data have
improved.  There are also theoretical reasons to suspect that
significant numbers of even lower surface brightness dwarfs could
exist \citep[e.g.,][]{bullock10}.  In the next decade, LSST
observations will reveal whether there is a large population of even
feebler dwarf galaxies, or if we have already reached the lowest
surface brightness at which galaxies are able to form.

\section{STELLAR POPULATIONS AND GAS CONTENT}
\label{sec:stellar_pops}

The gold standard for determining star formation histories based on
resolved stellar populations is \emph{Hubble Space Telescope} (HST)
photometry, because of its superior photometric accuracy and stability
relative to ground-based data.  One challenge facing such work for
UFDs is small-number statistics.  Obtaining strong constraints on the
star formation history of an old stellar population requires a sample
of at least $\sim200-300$ stars near the main sequence turnoff
\citep{brown14}.  The lowest-luminosity dwarfs simply do not contain
enough stars to meet this criterion even if every star in the galaxy
is observed.  Accurate star formation histories can be obtained for
systems with absolute magnitudes brighter than $M_{{\rm V}} \approx
-3$, although doing so may require a number of HST pointings in order
to include as many stars as possible.  HST-based star formation
histories have been published for 6 UFDs.

\subsection{Star Formation Histories}

The first analysis of deep HST imaging of UFDs was carried out by
\citet{brown12}, studying Hercules, Leo~IV, and UMa~I.  They concluded
that the three galaxies have similar ages, and are each as old or
older than the prototypical ancient globular cluster M92.
\citet{brown14} expanded the sample to six UFDs, adding Boo~I,
Canes~Venatici~II (CVn~II), and Coma~Berenices (Com~Ber) to the
previous three.  By incorporating improved spectroscopic MDFs and
updated isochrones matched to observed dwarf galaxy chemical abundance
patterns, \citet{brown14} determined that all of the galaxies except
UMa~I had formed more than 75\% of their stars by $z \sim 10$.  Using
a star formation model consisting of two bursts, the best fit for
UMa~I has approximately half of its stars forming at $z \sim 3$.  A
large majority of the stars in all six dwarfs had formed by the end of
reionization at $z \sim 6$, consistent with the idea that gas heating
by reionization ended star formation in such objects
\citep[e.g.,][]{bullock00,somerville02,benson02}.  Note, however, that
quenching by reionization does not necessarily mean that star
formation ends precisely at the redshift of reionization, since
sufficiently high-density molecular gas can survive somewhat beyond
reionization even in low-mass halos \citep[e.g.,][]{onorbe15}.  Star
formation histories have also been derived for Hercules, Leo~IV, and
CVn~II by \citet{weisz14a} from shallower WFPC2 data.
\citeauthor{weisz14a} found that $>90$\% of the stars in Hercules and
Leo~IV are older than 11~Gyr, consistent with the \citet{brown14}
results.  In CVn~II, however, \citeauthor{weisz14a} concluded that
star formation continued until $\sim8$~Gyr ago, in conflict with
\citeauthor{brown14} The reason for this discrepancy is not clear.
Age estimates based on deep ground-based imaging are generally
consistent with the HST results, although the constraints are not as
tight \citep[e.g.,][]{sand10,okamoto12}.

Based on the available data, it appears likely that UFDs are uniformly
ancient, with all or nearly all of their stars forming in the early
universe.  While most or all UFDs exhibit a blue plume of stars
brighter than the main sequence turnoff, this population is best
interpreted as blue stragglers rather than young stars
\citep{santana13}.  These objects can thus be considered pristine
fossils from the era of reionization \citep[e.g.,][]{br09,br11a,sf09}.
Improved age measurements to reveal how synchronized the star
formation in such galaxies was would be very interesting.  Conversely,
a clear detection of younger stars in very low-luminosity dwarfs would
have important implications for star formation in low-mass dark matter
halos and perhaps for cosmology as well \citep[e.g.,][]{bozek18}.

\subsection{Initial Mass Functions}

Their low metallicities make UFDs some of the most extreme
environments in which star formation is known to have occurred.  They
therefore present an promising opportunity to investigate how the
stellar initial mass function (IMF) depends on galactic environment.
Dwarf galaxies also offer the advantage that their low stellar
densities mean that no dynamical evolution has occurred, unlike in
globular clusters, so the present-day mass function can be assumed to
match the initial one below the main sequence turnoff.  \citet{geha13}
measured the IMF in two ultra-faint dwarfs, Hercules and Leo~IV, using
star counts from the HST photometry of \citet{brown12}.  Over the mass
range from $\sim0.5-0.8$~\msun, they found that the best fitting power
law had a slope of $\alpha \approx 1.2$, much shallower than the
\citet{salpeter55} value of $\alpha = 2.35$.  While the uncertainties
for Leo~IV are quite large, in Hercules the slope disagrees with a
Salpeter IMF at $5.8\sigma$.

Intriguingly, such a bottom-light IMF in the least massive, lowest
metallicity galaxies known suggests the possibility of a monotonic
trend in IMF slope with galaxy properties.  The largest elliptical
galaxies have bottom-heavy IMFs \citep[e.g.,][]{vdc10,spiniello12},
and dwarf galaxies in the Local Group appear to exhibit increasingly
shallower IMF slopes toward lower masses \citep{geha13}.

More recent analyses have complicated this picture.
\citet{gennaro18a} measured the IMFs for the full sample of 6 UFDs
from \citet{brown14}, confirming that each galaxy has an IMF slope
shallower than Salpeter when fit with a power law.  However, when
describing the IMF as a log-normal function \citep{chabrier03}, the
parameters for the UFDs are consistent with the Milky Way IMF.
\citet{gennaro18b} used deeper, near-infrared imaging of Com~Ber with
HST to probe the IMF down to masses of $\sim0.2$~\msun, comparable to
the characteristic mass in the log-normal description of the Milky Way
IMF.  The results are consistent both with the shallower optical
observations of Com~Ber and the \citet{chabrier03} Galactic IMF.
These findings suggest that there may be significant IMF variations
even within the class of UFDs, with some galaxies having shallow IMFs
and others that resemble the Milky Way despite their low metallicities
\citep{gennaro18b}.

If the IMF is indeed bottom-light in UFDs, there would be important
implications for SN rates, feedback, chemical enrichment, and gas loss
in such systems.  A bottom-light IMF extrapolated to higher masses is
top-heavy, which would produce larger numbers of SN explosions for a
given mass of stars (of course, the validity of such an extrapolation
is only an assumption, since no stars heavier than
$\sim0.8$~\msun\ exist in UFDs today).  This effect can be dramatic;
\citet{fsk14} estimated that for Segue~1 (with a present-day stellar
mass of $\sim500$~\msun), the galaxy would have hosted $\sim15$
core-collapse SNe for a Salpeter IMF compared to $\gtrsim250$ SNe for
the \citet{geha13} IMF.  Until the behavior of the IMF in the
ultra-faint dwarf regime is better understood, the number of SNe
expected to have occurred in such systems will be highly uncertain.

\subsection{Gas Content}

Among the dwarfs discovered since the beginning of SDSS, only Leo~T
(which we do not consider a UFD; see Section~\ref{sec:definition})
contains any neutral gas \citep{irwin07,rw08}.  Stringent upper limits
have been placed on the \hi\ content of many of the UFDs using
archival data or deep pointed observations with large single-dish
telescopes \citep{gp09,spekkens14,westmeier15}.  For the most nearby
dwarfs these limits can be as small as $\sim100$~\msun, while for
objects at distances of $\sim100$~kpc typical limits are
$\sim1000$~\msun.  No ionized gas associated with UFDs has been
detected either, but searches for low surface brightness H$\alpha$
emission similar to what has been detected for high-velocity clouds
\citep[e.g.,][]{putman03,barger12} could be of interest.

The lack of gas in these tiny galaxies is not a surprise, but the
mechanism by which they lost their gas is not clear.  Plausible
hypotheses for gas removal include reionization, supernova feedback,
and ram-pressure stripping.  Because nearly all currently known UFDs
are close to massive galaxies that are likely surrounded by hot
gaseous halos, ram-pressure stripping cannot be ruled out.  Studies of
isolated UFDs, which should be discovered with LSST, may shed light on
this issue.

\section{THE ULTRA-FAINT END OF THE GALAXY LUMINOSITY FUNCTION}
\label{sec:lf}

One of the key properties of the population of UFDs, as distinct from
the properties of individual objects, is their luminosity function
(LF).  The relationship between the LF and the mass function of dark
matter halos and subhalos encodes the physics of galaxy formation in
the smallest halos and places constraints on dark matter models.
Moreover, the LF provides the connection between the low luminosity
galaxies observed today and their progenitor systems at high redshift,
which may play a significant role in reionizing the universe (see
Section~\ref{sec:highz}).

The observed dwarf galaxy LF is only equal to the true LF if the dwarf
galaxy sample is complete over the luminosity range of interest.
UFDs, however, vary widely in luminosity, surface brightness, and
distance, and many are close to the detection limits of the surveys in
which they were discovered.  The LF of such systems therefore cannot
be computed until the sensitivity of dwarf galaxy searches has been
accurately quantified.

\citet{koposov08} presented a careful analysis of the detectability of
faint dwarf galaxies using an automated search algorithm in the 5th
data release (DR5) of SDSS, covering 8000~deg$^{2}$.  They found that
a significant fraction of the UFDs discovered in SDSS are close to the
detection limit of their algorithm.  These objects are detected in
SDSS data with an efficiency of $\sim50$\%, indicating that undetected
dwarfs are likely to be present in the SDSS footprint.  After
correcting for incompleteness, \citeauthor{koposov08} determined that
the differential LF of Milky Way satellites can be approximated as
$dN/dM_{{\rm V}} = 10^{0.1(M_{{\rm V}}+5) + 1}$ over the absolute
magnitude range $-19 < M_{{\rm V}} < -2$.  Translated into the
Schechter form, the corresponding faint-end slope of the LF is $\alpha
= -1.25$.  The implied total number of satellite galaxies within the
virial radius of the Milky Way is 45 at $M_{{\rm V}} < -5$ and 85 at
$M_{{\rm V}} < -2$.  For the faintest dwarfs the incompleteness
correction is very large, and it depends on the assumed radial
distribution of satellites.  If faint dwarfs are concentrated close to
the Milky Way, then fewer such objects are expected at large distances
where they are currently undetectable.  Conversely, if the spatial
distribution of the lowest luminosity systems is more extended then
there may be enormous numbers of similar objects in the outer halo of
the Galaxy.  While the radial distribution of dwarfs around the Milky
Way can be estimated in numerical simulations
\citep[e.g.,][]{wang13,gk17b}, ultimately it will have to be measured
observationally by deeper surveys.

A similar quantification of dwarf galaxy detectability was carried out
by \citet{walsh09} on SDSS DR6 imaging, covering 9500~deg$^{2}$.
\citeauthor{walsh09} used a more sensitive search algorithm that finds
all of the SDSS satellites known at the time at $\gtrsim90$\%
efficiency, but potentially with a correspondingly high false positive
rate.  They concluded that the transition between detectability and
invisibility as a function of luminosity, surface brightness, and
distance is more gradual than calculated by \citet{koposov08}, and
therefore that all of the known dwarfs should have been visible in
SDSS even if they were located at significantly larger distances.
According to the \citet{walsh09} analysis, searches in the SDSS
footprint are complete out to the virial radius of the Milky Way down
to $M_{{\rm V}} = -6.5$.  The extrapolated total number of Milky Way
satellites is $\sim220-340$ depending on the adopted detection
threshold.  Many subsequent studies have used the detection
sensitivity derived by \citet{koposov08} and/or \citet{walsh09} to
estimate the overall size of the Milky Way satellite population,
generally predicting that future surveys will discover $\sim100-300$
dwarfs over the entire sky
\citep[e.g.,][]{tollerud08,hargis14,newton18,jethwa18}.

Unfortunately, no comparable analyses of the detectability of dwarf
galaxies have been published since SDSS DR6.  Consequently, the
sensitivity of the final 5000~deg$^{2}$ of SDSS imaging, the Dark
Energy Survey, Pan-STARRS, and other smaller surveys has never been
adequately quantified.  Given the large number of new satellites
discovered since 2009 and their apparently anisotropic distribution on
the sky \citep{drlica15}, updated determinations of the completeness
of searches for nearby dwarf galaxies are urgently needed.  Until the
sensitivity of all significant surveys has been properly quantified,
more detailed calculations of the total number of Milky Way
satellites, their LF, and their radial and angular distributions
cannot be made.  What we can say at present is that the observed LF
(without an incompleteness correction) peaks at $M_{{\rm V}} \sim -4$,
suggesting that any real turnover in the LF must be at even fainter
magnitudes.

As an illustration of the discovery potential of future imaging
surveys, we construct a very simple toy model of satellite
detectability.  Motivated by the results of \citet{martin08} for SDSS
and \citet{bechtol15} and \citet{drlica15} for DES, we assume that a
satellite must contain at least 20 stars brighter than the detection
limit of the survey in order to be identified.  We create realizations
of satellites with stellar masses corresponding to absolute magnitudes
of $M_{{\rm V}} = -2, -4$, and $-6$ by randomly selecting the
appropriate number of stars from an old, metal-poor mock stellar
population.  We then determine the median magnitude of the 20th
brightest star for each absolute magnitude and calculate out to what
distance that star would be detectable for a given survey depth.  Note
that the depth of a survey for the purpose of searching for stellar
overdensities is $\gtrsim0.5$~mag shallower than the actual $5\sigma$
detection limit because colors become uncertain and star-galaxy
separation becomes unreliable at fainter magnitudes.  Consistent with
\citet{koposov08} and \citet{walsh09}, we find that SDSS should be
complete at $M_{{\rm V}} \approx -6$ out to beyond the virial radius
of the Milky Way.  Similarly, DES should be complete down to $M_{{\rm
    V}} \approx -4$ within the virial radius.  A complete search of
the Milky Way's virial volume to fainter magnitudes will require
full-depth LSST images.

\begin{figure}[h]
\includegraphics[width=6.33in]{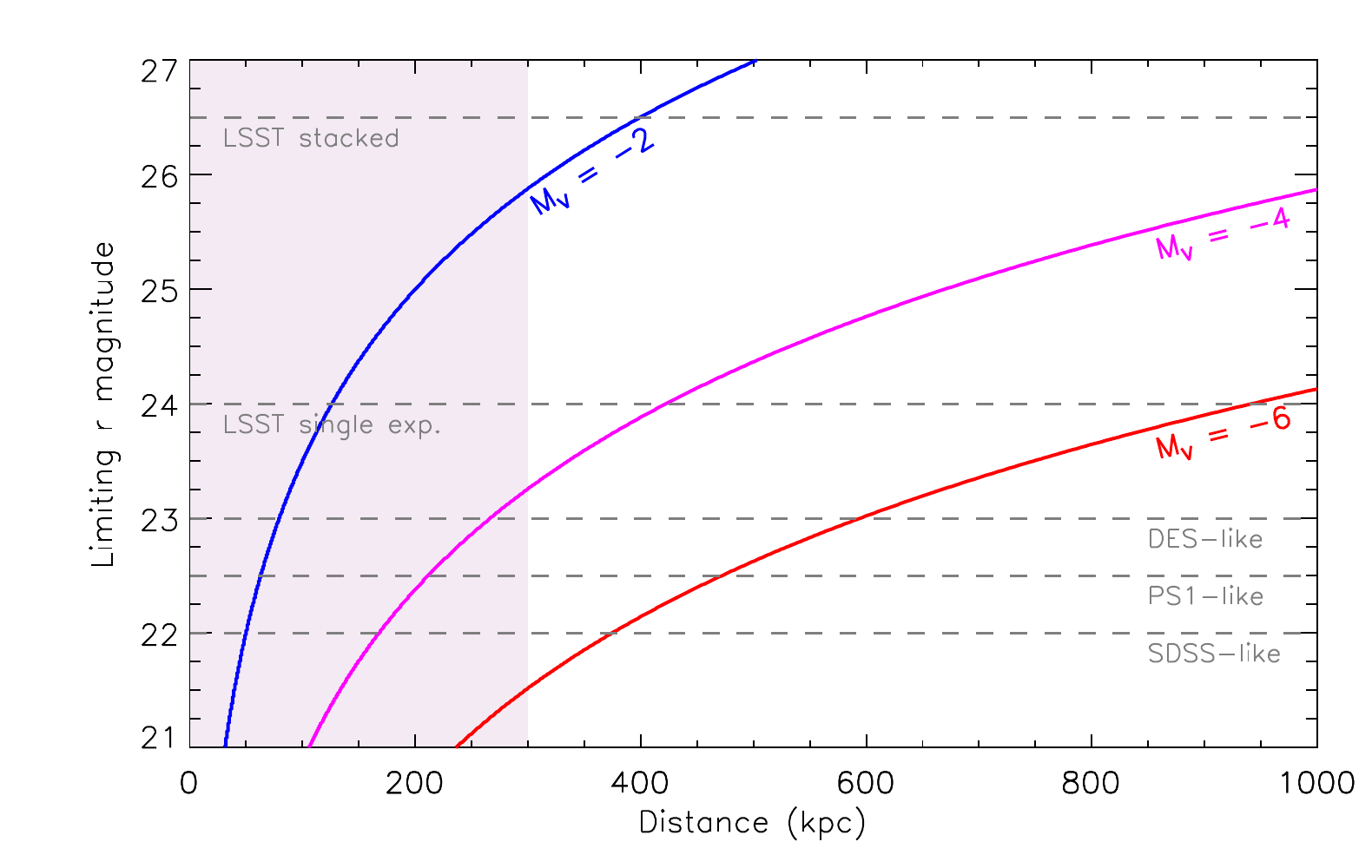}
\caption{Detectability of faint stellar systems as a function of
  distance, absolute magnitude, and survey depth.  The red curve shows
  the brightness of the 20th brightest star in an $M_{{\rm V}} = -6$
  object as a function of distance.  The magenta and blue curves show
  the brightness of the 20th brightest stars for $M_{{\rm V}} = -4$
  and $M_{{\rm V}} = -2$ systems, respectively.  The horizontal dashed
  lines indicate (from bottom to top) the limiting r magnitude for
  dwarf galaxy searches in SDSS, Pan-STARRS, DES, LSST single
  exposures, and stacked LSST images at the end of the survey.  The
  region within the (approximate) virial radius of the Milky Way is
  shaded purple.}
\label{fig:sensitivity}
\end{figure}

\section{ORIGIN AND EVOLUTION}
\label{sec:origin}

\subsection{The Formation of Ultra-Faint Dwarfs and the Stellar
  Mass-Halo Mass Relation}

The formation of the first galaxies depends critically on the
mechanisms that allow gas to cool to low enough temperatures for star
formation to begin (see, e.g., \citealt{by11} and references therein).
UFDs may form either in dark matter minihalos of $10^{6} -
10^{8}$~\msun\ \citep[e.g.,][]{br09,sf09}, which cool via molecular
hydrogen and are thought to be the hosts of the first Population III
stars at $z\sim20$, or in atomic cooling halos of $> 10^{8}$~\msun,
which cool initially via atomic hydrogen lines and collapse later at
$z\sim10$ \citep[e.g.,][]{li10,fb12}.  Observationally, it may be
possible to distinguish these scenarios via either the present-day
halo masses of UFDs or their chemical enrichment.

Simulating such tiny galaxies is a difficult computational problem
because of the high resolution and high dynamic range needed.  There
are at least three approaches used in the literature to study small
dwarf galaxies theoretically.  First, one can directly carry out
ultra-high resolution zoom-in simulations of the first galaxies, which
explore the physics of the formation and evolution of such systems,
but the simulations are generally too expensive to run to the present
day \citep[e.g.,][]{wise14,jeon15}.  Alternatively, simulations of
dwarf galaxies located in isolated environments can be run to $z=0$,
at the cost of missing the physics associated with satellite dynamics
and stripping \citep[e.g.,][]{simpson13,wheeler15}.  Finally,
simulations of satellites of Milky Way-like galaxies may include all
of the relevant physics and run to $z=0$, but are only just beginning
to reach the resolution required to study UFDs
\citep[e.g.,][]{wetzel16, gk18}.  In the latter two classes of
simulations the properties of the simulated dwarfs are typically in
reasonable agreement with observations
\citep[e.g.,][]{wetzel16,jeon17}, but more computing power and higher
resolution will be needed to investigate the formation of ultra-faint
dwarfs in detail.

One of the most basic questions regarding the formation of dwarf
galaxies in a cosmological context is what dark matter halos they
occupy.  The properties of their dark matter halos control when dwarfs
form, their gas content, and their resilience to heating by stellar
feedback and reionization.  The correspondence between galaxy stellar
masses and dark matter halo masses is referred to as the stellar
mass-halo mass (SMHM) relation.  As emphasized by \citet{bp17} and
\citet{kim18}, the SMHM relation is the key to understanding most of
the so-called small scale challenges to the $\Lambda$CDM model
\citep[e.g.,][]{bbk17}.

For halo masses below $\sim10^{12}$~\msun, the SMHM relation is
generally described as a power law, $M_{*} \propto M_{{\rm
    halo}}^{\alpha}$ with small scatter \citep{moster13,behroozi13}.
How the relation behaves for stellar masses below $\sim10^{7}$~\msun
(i.e., for classical and ultra-faint dwarfs) is currently a matter of
debate.  Hydrodynamic simulations suggest very large scatter at these
lowest masses, with an increasing fraction of halos remaining
completely dark \citep[e.g.,][]{shen14,sawala16b,munshi17}.  However,
\citet{jethwa18} argued based on Bayesian fits to the Milky Way
satellite LF with a wide variety of SMHM parameterizations that the
observational data are matched better without large scatter, and that
the fraction of halos hosting observable galaxies must be significant
even at quite low masses.  Alternatively, \citet{re18} recently
proposed that a relation between the mean star formation rate and halo
mass may be better constrained at dwarf galaxy masses than the SMHM
relation is.  The discovery of additional satellites and completeness
analyses of surveys beyond SDSS will offer improved constraints on the
matching between halos and UFDs, which may provide the solution to the
missing satellite problem \citep{kim18,re18}.

\subsection{Galactic Orbits}

The orbits of dwarf galaxies around the Milky Way control both their
tidal evolution and potentially their gas loss through ram-pressure
stripping or other effects.  However, without three-dimensional
kinematic information, only weak orbital constraints are possible.  By
comparing radial velocity measurements with the Via Lactea II N-body
simulation, \citet{rocha12} were able to determine approximately when
the classical and SDSS satellites last crossed the virial radius of
the Milky Way, but the specific orbits of each dwarf remained unknown.
For the 15 UFD candidates that lack radial velocities
(Section~\ref{sec:kin}), even the infall times cannot be measured.

Until very recently, the only published proper motion for a UFD was
the ground-based measurement for Segue~1 (the closest dwarf) by
\citet{fritz18}.  However, the situation changed dramatically as soon
as the second data release (DR2) from \emph{Gaia} became available.
\citet{gaiadr2helmi} measured the proper motion and orbit of Boo~I,
and \citet{simon18}, \citet{fritz18b}, and \citet{kallivayalil18}
immediately determined the proper motions and orbits of all UFDs for
which spectroscopic members were available.  At magnitude $G = 16, 17,
18,$ and 19, \emph{Gaia} DR2 provides typical proper motion
uncertainties of 0.09, 0.15, 0.26, and 0.51~mas~yr$^{-1}$,
respectively \citep{lindegren18}, so even a single UFD member star
brighter than $G \approx 18$ is sufficient to determine an accurate
proper motion for a given galaxy.  The corresponding tangential
velocity uncertainties are $43, 71, 123,$ and $241 (d/100{\rm
  kpc})$~\kms, where $d$ is the distance in kpc.  By averaging the
proper motions of multiple members, smaller uncertainties can be
achieved.  Remarkably, the \emph{Gaia} astrometry is so accurate that
UFD proper motions can be measured with a combined photometric and
astrometric selection even without any spectroscopic membership
information \citep{mh18,pl18}.  We are therefore suddenly in the
situation where UFD proper motions outnumber radial velocities.  For
UFDs beyond $\sim150$~kpc, the \emph{Gaia} DR2 proper motions are
generally consistent with zero, but that can be improved by the
identification of additional member stars and by future \emph{Gaia}
releases.

Orbits based on full six-dimensional phase space information have now
been computed for all of the Milky Way's ultra-faint satellites with
known radial velocities
\citep{gaiadr2helmi,simon18,fritz18b,kallivayalil18}.  The most
important orbital parameter is the pericenter distance, which
fortunately only depends rather weakly on the assumed gravitational
potential of the Milky Way.  For the satellites within 100~kpc, the
median pericenter is 38~kpc, the pericenter is generally determined to
within $10-20$\%, and the typical orbital period is several Gyr
\citep{simon18,fritz18}.  Surprisingly, this orbit modeling reveals
that nearly all of the closest dwarfs are currently very close to the
pericenters of their orbits.  Only Boo~I, Willman~1, and Tucana~II are
more than $\sim5$~kpc beyond their pericenter distances
\citep{simon18}.  The most natural explanation for this peculiar
positioning is that there is a selection bias against discovering UFDs
that are far from their orbital pericenters.  If that is the case,
then most of the dwarfs found in SDSS, DES, and other surveys must be
close to the survey detection limits, as suggested by
\citet{koposov08}, but contrary to the results of \citet{walsh09},
\citet{bechtol15}, and \citet{drlica15}.  Deeper surveys should then
reveal a significantly larger population of UFDs that are distributed
more evenly along their orbits.

The satellites discovered in DES imaging are noticeably concentrated
around the Large and Small Magellanic Clouds
\citep{bechtol15,drlica15}.  This result led to speculation that many
of these objects might have originated as Magellanic satellites and
are now being accreted by the Milky Way
\citep{deason15,drlica15,jethwa16,sales17}, as originally predicted by
\citet{dl08}.  Based on their space motions and orbits, the most
likely dwarfs to have formed in the Magellanic group are Horologium~1,
Hyi~I, Carina~III, and Tucana~II \citep{kallivayalil18,simon18}.

\subsection{Tidal Evolution}

Since their discovery, it has frequently been suggested that many UFDs
are experiencing significant tidal stripping as they orbit the Milky
Way
\citep[e.g.,][]{belokurov06,zucker06b,no09,munoz10,sand12,kirby13,roderick15,collins17,simon17}.
The objects upon which this speculation has focused include Hercules,
Leo~V, UMa~I, UMa~II, Segue~1, Segue~2, and Tuc~III.  The physical
reasoning supporting the idea of tidal stripping or tidal disruption
for these satellites ranges from morphology (highly elongated shapes;
Hercules and UMa~I), apparent extratidal features (Hercules, Segue~1,
and Tuc~III), and possible velocity gradients (Hercules and Leo~V) to
deviations from the luminosity-metallicity relation (Segue~2).

Now that the orbits of the UFDs are known, the possibility of tidal
stripping or tidal disruption can be discussed much more
quantitatively.  While some stripping of the dark matter from
satellite galaxies is inevitable on almost any orbit, they must
approach the Milky Way much more closely in order for an appreciable
fraction of their stars to be lost \citep{penarrubia08}.  The tidal
radius of a dwarf galaxy is unfortunately not a well-defined quantity,
as it varies with time and depends on both the poorly-known mass
distribution of the Milky Way and the even more poorly-known mass
distribution of the dwarf.  In lieu of carrying out detailed numerical
experiments for each dwarf, one can approximate the tidal radius as
the Jacobi radius \citep{bt08}:

\begin{equation}
r_{t} = \left( \frac{m_{{\rm dwarf}}}{3M_{{\rm MW}}} \right)^{1/3} d,
\end{equation}

\noindent
where $m_{{\rm dwarf}}$ is the mass of the dwarf galaxy, $M_{{\rm
    MW}}$ is the mass of the Milky Way interior to the location of the
dwarf, and $d$ is the distance of the dwarf from the Galactic center.
However, we also encourage future computational work to model the
response of stars within a dwarf to a time-variable external tidal
field in a more realistic way.

To conservatively assess stripping, we adopt a heavy ($1.6 \times
10^{12}$~\msun) Milky Way model and assume that the total mass of each
dwarf is limited to the measured mass within its half-light radius.
Since the actual halo mass of a dwarf is expected to be several orders
of magnitude larger in the absence of stripping, this scenario places
a lower bound on the tidal radius.  Under these assumptions, we
calculate that the tidal radius is currently beyond $3 R_{1/2}$ for
all of the UFDs except Tuc~III ($r_{t} = 2.3 R_{1/2}$) and UMa~I
($r_{t} = 2.2 R_{1/2}$).  The former is not surprising, as Tuc~III has
an orbital pericenter of only $\sim3$~kpc
\citep{erkal18,simon18,fritz18}, at which point its tidal radius may
be smaller than its present half-light radius.  Not coincidentally,
Tuc~III is also the only UFD that is unambiguously suffering
substantial stripping, with clear tidal tails comprising the majority
of its stellar mass extending at least 1~kpc from its main body
\citep{drlica15,shipp18,li18b,erkal18}.  UMa~I, on the other hand, has
a pericenter of $\approx101$~kpc \citep{simon18,fritz18}, essentially
equal to its current distance.  Its tidal radius is therefore at its
minimum value now, and the outer $\sim15-20$\% of its stars may be
vulnerable to stripping.  Recall, however, that we have made the
extreme assumption that the dark matter halo of each dwarf is
truncated at its half-light radius.  If the halo of UMa~I is
substantially more extended, as is very likely to be the case, then
only minimal stripping of its stars is possible.

For the remaining UFDs with published kinematics, significant
stripping generally appears unlikely.  At the pericenters of their
orbits, Hyi~I and Boo~I each have $r_{t} \approx 3 R_{1/2}$, which
would leave $\approx10$\% of their stars unbound.  Again, though, more
realistic assumptions about the mass and extent of their dark halos
would result in no significant stellar stripping.  Segue~2 is in
danger of stripping if its velocity dispersion is much smaller than
the upper limit determined by \citet{kirby13}.  If $\sigma \lesssim
0.7$~\kms, then its tidal radius would be $\approx 2 R_{1/2}$ at
pericenter, so tides remain a plausible explanation for its offset
from the luminosity-metallicity relation.  The more distant dwarfs
often regarded as likely to have been stripped or disrupted, Hercules
and Leo~V, can only experience tidal stripping if they are on
extremely eccentric orbits bringing them within $10-20$~kpc of the
Galactic center.  Such an orbit is not currently excluded for
Hercules, but is unlikely for Leo~V \citep{fritz18}.  With larger halo
masses these dwarfs would need to pass within a few kpc of the Milky
Way to be disrupted.  We therefore suggest that alternative
explanations for elongated shapes and velocity gradients, such as
formation through mergers or puffy disks
\citep[e.g.,][]{starkenburg16,wheeler17}, should be considered before
necessarily attributing such properties to Milky Way tides.

\section{ULTRA-FAINT DWARFS AS DARK MATTER LABORATORIES}
\label{sec:dm}

The nature of dark matter is one of the most significant outstanding
questions in astrophysics, and the smallest dwarfs may play an
outsized role in helping to answer it.  In this section we mention
some of the ways in which UFDs can constrain dark matter properties
and dark matter models.  For broader discussions of dwarf galaxies
from a dark matter perspective, see, e.g., \citet{porter13},
\citet{weinberg15}, \citet{bbk17}, \citet{bp17}, or
\citet{strigari18}.

UFDs can potentially provide insight into dark matter for several
reasons:

\begin{itemize}
\item They are the most dark matter-dominated systems known.  Unlike
  in larger and more luminous dwarfs \citep[e.g.,][]{bz14,dicintio14},
  their baryonic components are likely to have been dynamically
  negligible at all times.  Their inefficient star formation means
  that feedback should not be powerful enough to alter their internal
  density structure \citep[e.g.,][]{onorbe15}.

\item Because of their small sizes, they offer probes of dark matter
  on smaller scales ($\sim20-30$~pc for the most compact ultra-faint
  dwarfs) than is possible anywhere else.

\item The number of dwarf galaxies orbiting the Milky Way sets a
  lower bound on the abundance of low-mass dark matter subhalos, which
  translates to a limit on the allowed mass of warm dark matter
  particles \citep[e.g.,][]{kennedy14}.

\item With the exception of the Galactic Center and Sagittarius (which
  they greatly outnumber), they are the closest dark matter halos to
  us.  The combination of their proximity, their high measured
  densities \citep[e.g.,][]{sg07}, and their low astrophysical
  backgrounds makes them promising targets for indirect detection
  experiments.

\item Their internal dynamics are so gentle that heating by very weak
  effects is potentially measurable.  For example, \citet{brandt16}
  used the presence of a star cluster in Eridanus~II to place tight
  constraints on MACHO dark matter, and \citet{penarrubia16} proposed
  that wide binary stars may be disrupted by the dark matter potential
  of a UFD, allowing a measurement of the dark matter density profile.

\end{itemize}

Because of the arguments listed above, UFDs have attracted a great
deal of attention from a broad cross-section of astrophysicists.
Their potential to facilitate indirect detection of dark matter has
been a particular focus of attention.  The majority of indirect
detection experiments search for gamma-rays resulting from
annihilation of dark matter particles, using either the \emph{Fermi
  Gamma-Ray Space Telescope} or ground-based atmospheric Cherenkov
telescopes.  UFDs are prime targets for both types of facilities
\citep[e.g.,][]{magic16,albert17,archambault17}.  The sensitivity of
these searches will continue to improve as integration times increase
and new observatories such as the Cherenkov Telescope Array begin
operation.  Indirect detection searches in UFDs will offer a critical
testing ground for possible dark matter signals seen in other parts of
the sky \citep[e.g.,][]{ak16}.  Dark matter annihilation or decay
signals could also manifest in dwarf galaxies as synchrotron emission
at radio wavelengths \citep{spekkens13,regis17} or as X-ray emission
lines \citep[e.g.,][]{jp16}.

The holy grail for dark matter research in dwarf galaxies is the
conclusive measurement of the inner density profile of a highly dark
matter-dominated system.  As mentioned above, UFDs are ideal in the
sense that they have the highest known dark matter fractions of any
galaxies, and their density structure is unlikely to have been
affected by stellar feedback.  Their disadvantage is that they contain
so few stars that there may not be enough dynamical tracers for a
robust measurement of the mass distribution.  Given the difficulties
encountered in analyzing radial velocity data sets containing hundreds
to thousands of stars in the classical dSphs, the maximum achievable
sample of $\sim100$ stars in the most accessible UFDs will not be
sufficient to separate a central dark matter cusp from a core.
However, the combination of radial velocities and proper motions can
provide much more accurate measurements
\citep[e.g.,][]{strigari07,kallivayalil15}.  Measuring proper motions
with an accuracy of $\sim35$~$\mu$as~yr$^{-1}$ (5~\kms\ at a distance
of 30~kpc) for stars as faint as $r \sim 22$ is a daunting task, but
may be feasible with extremely large ground-based telescopes or by
combining data from space-based facilities such as HST, \emph{Gaia},
JWST, and WFIRST.

\section{ULTRA-FAINT DWARFS BEYOND THE MILKY WAY}
\label{sec:beyond_mw}

\subsection{Ultra-Faint Dwarfs Around M31}

The natural first step in studying UFDs beyond the Milky Way is
exploring the vicinity of M31.  The Pan-Andromeda Archaeological
Survey (PAndAS) has now imaged the M31 halo out to a projected radius
of $\sim150$~kpc using the Canada-France-Hawaii Telescope
\citep{mcconnachie09}, discovering 17 new dwarf galaxies
\citep{martin06,martin09,martin16c,ibata07,irwin08,mcconnachie08,richardson11}.
An additional 8 dwarfs in the vicinity of Andromeda have also been
discovered since 2004, mainly in SDSS and Pan-STARRS
\citep{zucker04,zucker07,majewski07,slater11,bell11,martin13b,martin13c}.
While a handful of these dwarf galaxies may not be true satellites of
M31, all of them except Andromeda~XXVII \citep{conn12} are likely
located within the M31 virial radius.  The currently known M31
satellite population reaches as faint as $M_{{\rm V}} \approx -6$
\citep{martin16c}, including 8 UFDs according to our definition.  The
sizes and luminosities of the ultra-faint M31 satellites are in
excellent agreement with the locus established by Milky Way dwarfs in
Fig.~\ref{fig:mv_rhalf}.

\subsection{Surveys Outside the Local Group}

Detecting UFDs at even larger distances is difficult because of their
low surface brightnesses and small sizes.  In the nearest galaxy
groups at distances of $3-4$~Mpc, the most luminous red giants have
apparent magnitudes of $r \sim 24.5-25$.  Since faint dwarfs contain
few stars near the tip of the red giant branch, imaging to fainter
than 26th magnitude is necessary to identify an ultra-faint dwarf at
these distances as an overdensity of resolved stars.  In dedicated
deep surveys and HST imaging of nearby galaxy clusters, several
objects near or below our magnitude limit separating UFDs from dSphs
have recently been discovered, including d0944+69 \citep[$M_{{\rm V}}
  = -6.4$;][]{chiboucas09,chiboucas13} in the M81 group, Virgo~UFD1
\citep[$M_{{\rm V}} = -6.5$;][]{jl14} in the Virgo cluster,
CenA-MM-Dw7 \citep[$M_{{\rm V}} = -7.2$;][]{crnojevic16} in the
Centaurus~A group, MADCASH~J074238+652501-dw \citep[$M_{{\rm V}} =
  -7.7$][]{carlin16} around NGC~2403, and Fornax~UFD1 \citep[$M_{{\rm
      V}} = -7.6$;][]{lee17} in the Fornax cluster.  Low-surface
brightness dwarfs in the Local Volume with luminosities in the UFD
regime can also be identified via their diffuse light
\citep[e.g.,][]{bennet17,danieli18}.  The sample of UFDs in other
environments is still too small and heterogeneous for comparative
studies, but the luminosities and radii of these dwarfs seem to be
consistent with the properties of the Milky Way satellites shown in
Fig.~\ref{fig:mv_rhalf}.

The first significant sample of UFDs beyond the Local Group will
likely be revealed by LSST.  The stacked end-of-survey LSST images
will reach fainter than 27th magnitude in $g$ and $r$ bands, up to
$\sim1$~mag beyond the depth of the current state-of-the-art PISCeS
\citep[e.g.,][]{sand14,crnojevic16} and MADCASH \citep{carlin16}
surveys.  Extrapolating from current results, LSST should be sensitive
to galaxies as faint as $M_{{\rm V}} \approx -6$ in galaxy groups at
$3-4$~Mpc, and even lower luminosity systems in the local field at
$1-2$~Mpc \citep[e.g.,][]{tollerud08,lsstscibook} via resolved stars.
Systematic searches for UFDs throughout this volume will enable the
galaxy luminosity function to be probed down to extremely faint
absolute magnitudes across a wide range of environments.

In more massive dwarf galaxies ($M_{*} > 10^{7}$~\msun), population
studies demonstrate that star formation is shut off only by
environmental effects \citep{geha12}.  The lack of gas or ongoing star
formation among satellites of the Milky Way and M31 suggests that
starvation and ram-pressure stripping are the primary mechanisms for
environmental quenching down to masses as small as $M_{*} \approx
10^{5.5}$~\msun\ \citep{wetzel15,fillingham15,fillingham16,fillingham18}.
In the ultra-faint regime, however, the available star formation
histories show that star formation ended $\sim12$~Gyr ago
\citep{brown14} even though at least some of the galaxies were likely
accreted by the Milky Way more recently
\citep{rocha12,simon18,fritz18b}.  At lower stellar masses, the timing
of quenching, N-body-based models, and hydrodynamic simulations all
suggest that reionization is responsible for shutting off star
formation \citep{brown14,jeon17,fitts17,wimberly18}.  If this
hypothesis is correct, then UFDs can form anywhere and need not be in
close proximity to massive galaxies.  LSST would therefore be expected
to find large numbers of such systems beyond the boundary of the Local
Group \citep{wimberly18}.

\subsection{Connection to Observations of the High-Redshift Universe}
\label{sec:highz}

In a recent series of important papers, Boylan-Kolchin, Weisz, and
collaborators have quantified the correspondence between dwarf
galaxies observed today in the Local Group and faint galaxies at high
redshift.  \citet{bk15} used the observed star formation histories of
nearby dwarfs to calculate their ultraviolet (UV) luminosities as a
function of time.\footnote{At $z = 7$, $M_{{\rm UV}} = 0.71 M_{{\rm
      V}} (z=0) - 2.71$, such that classical dSphs had UV magnitudes
  in the reionization era similar to their V-band magnitudes today,
  while UFDs had high-$z$ UV magnitudes $\sim1-2$~mag brighter than
  their present-day optical magnitudes.}  They showed that reionizing
the universe require a significant contribution of UV photons from
galaxies at least as faint as the Fornax dSph.  Even with the
\emph{James Webb Space Telescope} such galaxies will not be detectable
at $z \sim 7$ \citep{bk15}.  Moreover, \citet{bk16} demonstrated via
comparison to N-body simulations that the Local Group is comparable in
size to the \emph{Hubble} Ultra Deep Field, and is a cosmologically
representative volume at dwarf galaxy masses.  \citet{wbk17} then
examined the UV LF in the reionization era.  Given that the observed
properties of UFDs today demonstrate that galaxies as faint as
$M_{{\rm UV}} \sim -3$ existed at high redshift, they showed that if
the currently measured faint-end slope of the UV LF ($\alpha \sim -2$;
e.g., \citealt{livermore17}) is extrapolated to $M_{{\rm UV}} = -3$
then UFDs dominate the ionizing photon production of the universe.
However, this assumption substantially overpredicts the observed dwarf
galaxy population of the Local Group.  If the faint-end slope is
shallower ($\alpha = -1.25$), as estimated by \citet{koposov08} from
SDSS data, then only bright dwarfs contribute to reionization.

This analysis highlights the complementarity between direct
observations of the epoch of reionization and studies of the ancient
stars in the closest galaxies.  Local Group observations can probe the
population of typical galaxies orders of magnitude fainter than will
be possible at high redshift in the foreseeable future.  As described
in the preceding sections, these galaxies can also be dissected star
by star, with detailed kinematic, chemical, mass, age, and spatial
information.  On the other hand, the distant universe provides much
better statistics, access to a variety of environments, and the
opportunity to compare galaxy populations across cosmic time, none of
which can be done nearby.  At the intersection between the two we can
learn about the sources the reionized the universe, the halo masses
associated with faint galaxies, and stellar populations and
nucleosynthesis in the first galaxies.

\section{SUMMARY AND OUTLOOK}
\label{sec:summary}

Our understanding of the faintest dwarf galaxies has progressed
rapidly since their discovery 14 years ago.  As described in
Sections~\ref{sec:intro} and \ref{sec:kin}, even the basic nature of
the first ultra-faint dwarfs was unclear for several years.  Now,
thanks to dedicated follow-up efforts across a wide range of
facilities, the velocity dispersions, masses, densities,
metallicities, metallicity dispersions, ages, IMFs, proper motions,
and orbits of subsets of the known UFDs have been measured.  These
observations have shown that UFDs are the most dark matter-dominated,
oldest, most metal-poor, and most chemically primitive stellar systems
known.  Concordant theoretical efforts devoted to simulating galaxy
formation in low-mass dark matter halos at increasingly high
resolution indicate that the faintest dwarfs appear to naturally
correspond to the luminous counterparts of the smallest halos capable
of sustaining star formation.

\begin{figure}[h]
\includegraphics[width=6.33in]{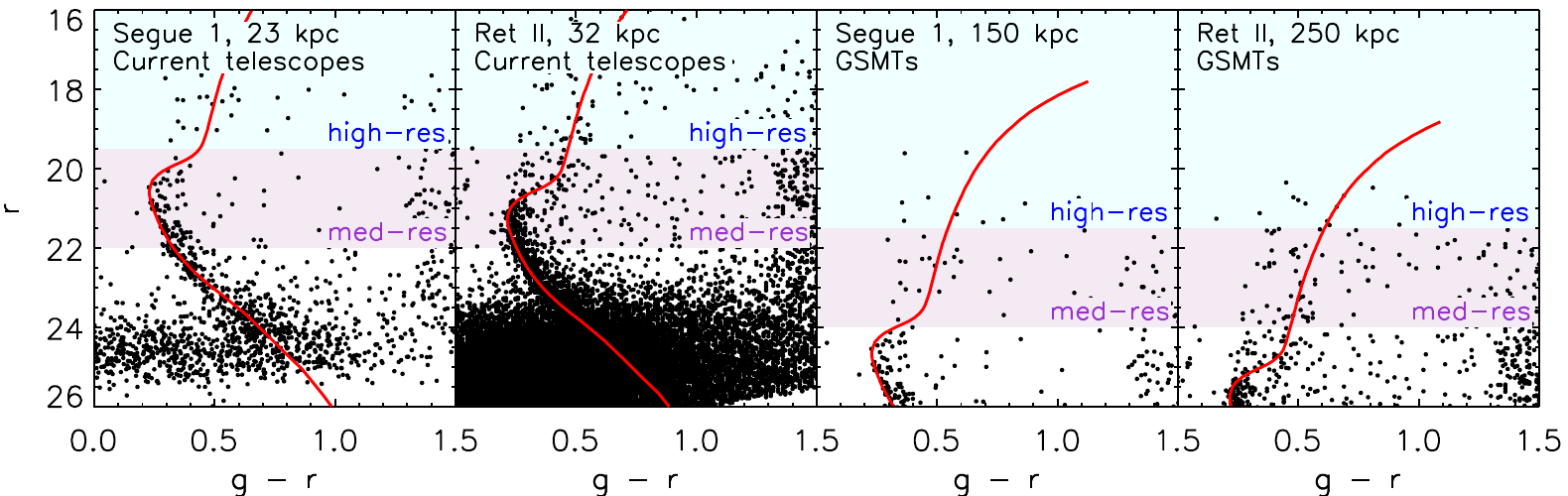}
\caption{(\emph{left panel}) Color-magnitude diagram of Segue~1
  (photometry from \citealt{munoz18b}).  The shaded blue and purple
  magnitude regions indicate the approximate depth that can be reached
  with existing medium-resolution and high-resolution spectrographs,
  respectively.  (\emph{left middle panel}) Same, but for Ret~II
  (using DES DR1 photometry).  (\emph{right middle panel}) Segue~1
  shifted to a distance of 150~kpc.  With current telescopes only a
  handful of its stars would be spectroscopically accessible.  The
  shaded blue and purple regions now indicate the depth that could be
  reached with 30-m telescopes.  (\emph{right panel}) Ret~II shifted
  to a distance of 250~kpc, again with the magnitude limits for 30-m
  telescope spectroscopy.}
\label{fig:cmd_speclimits}
\end{figure}

Much work remains to be done, of course.  No spectroscopy has been
obtained for $\sim1/3$ of the current ultra-faint satellite
population, leaving the status of some objects in question, and the
highest-quality star formation histories are available for only 6
galaxies.  On the theoretical side, simulating the formation and
evolution of low-mass dwarfs around a Milky Way-like host to $z=0$
remains a computational challenge.  Analogs to the lowest-luminosity
galaxies ($M_{*} \lesssim 10^{3}$~\msun) have not yet been reliably
simulated.  Importantly, the census of Milky Way satellites remains
significantly incomplete.  Even in the most pessimistic predictions,
the Milky Way has approximately twice as many dwarf satellites as have
been found so far \citep[e.g.,][]{newton18}.  In optimistic scenarios,
the total population could be nearly an order of magnitude larger.
The missing nearby satellites may be revealed in the next few years by
ongoing surveys such as MagLiteS \citep{dw16} and the DESI Legacy
Imaging Surveys \citep{dey18}, but the more distant ones will require
deeper imaging (e.g., LSST).  Discovering, confirming, and
characterizing possibly hundreds of dwarf galaxy candidates will be a
very large undertaking for the worldwide community.  As an
illustration of this challenge, Figure~\ref{fig:cmd_speclimits} shows
color-magnitude diagrams of the very low-luminosity dwarfs Segue~1 ($d
= 23$~kpc; $M_{{\rm V}} = -1.3$) and Ret~II ($d = 32$~kpc; $M_{{\rm
    V}} = -4.0$), along with the approximate spectroscopic limits that
can be achieved at medium resolution (for velocities) and high
resolution (for chemical abundances) with current facilities.
Spectroscopy for comparable systems at much greater distances can only
be obtained with 30~m-class telescopes.

Looking forward, after completing the census of dwarf galaxies
surrounding the Milky Way, stellar kinematics measurements can be used
to determine their mass function for comparison with theoretical
predictions.  Continued chemical reconnaissance via high-resolution
spectroscopy may provide new clues to the additional site(s) of
r-process nucleosynthesis, and with luck could reveal the signatures
of Population III SN explosions.  Now that the formation environments
of UFDs can be traced by their orbits, precision
measurements including an expanded sample of ages from space-based
photometry will show how dwarfs that formed in the field or in the
Magellanic Group differ from those that have always been close to the
Milky Way.  Mass and density measurements will provide critical
sensitivity and targeting information for indirect detection
experiments.  Although a detection may seem unlikely given current
limits, any signal from dark matter would be of such importance that
the search must continue.  Finally, we may hope that astrometry or
other novel techniques make it possible to determine the dark matter
density profiles of the least-perturbed dark matter halos yet found.
This measurement would strongly constrain the properties of dark
matter.  Given the tremendous amount we have already learned from
studying UFDs, it would be fitting if the humblest
galaxies in the Universe provided the answer to one of its biggest
questions.

\begin{issues}[FUTURE ISSUES]
\begin{enumerate}
\item Completing the census of Milky Way satellites.  LSST will be
  needed in order to detect the faintest currently known dwarfs
  throughout the virial volume of the Milky Way.  However, even
  achieving all-sky coverage (outside the Galactic plane) at SDSS
  depth or deeper, coupled with well-quantified detection limits, will
  substantially advance our knowledge of the galaxy luminosity
  function at faint magnitudes and the likely size of the satellite
  population of our Galaxy.
\item Obtaining photometric and spectroscopic follow-up observations
  for as much of the ultra-faint satellite population as possible.
  These observations are essential for classifying compact,
  low-luminosity stellar systems and determining their dark matter
  content.  Metallicity and age measurements will enable us to
  reconstruct their formation and evolution.  Expanding the current
  small sample of galaxies with precise star formation histories is an
  especially high priority for understanding the effects of
  reionization and environment on star formation in the faintest
  dwarfs.  Detailed chemical abundance patterns of UFD stars are
  likely to provide new insight into nucleosynthesis in the early
  universe.
\item Improving numerical simulations of the smallest galaxies.  At
  present, the computational strategy is often to adjust the physics
  in simulations in order to reproduce the observed properties of
  dwarfs.  As resolution increases, it should be possible to move
  beyond this approach and learn about the earliest stages of
  formation of these systems, and how they evolve in the gravitational
  potential of the Milky Way.  A key result from future simulations
  will be determining how galaxies populate dark matter halos at
  masses below $M_{{\rm halo}} \sim 10^{9}$~\msun.
\item Testing dark matter physics.  In addition to placing a lower
  limit on the mass function of dark matter subhalos, a sample of
  hundreds of stellar radial velocities and proper motions or other
  novel ideas could yield tight constraints on the inner density
  structure of UFD dark matter halos.  These measurements would
  provide a critical test of the cold dark matter prediction that the
  density profiles of undisturbed dark matter halos should have
  central cusps.
\end{enumerate}
\end{issues}

\section*{DISCLOSURE STATEMENT}
The author is not aware of any affiliations, memberships, funding,
or financial holdings that might be perceived as affecting the
objectivity of this review.

\section*{ACKNOWLEDGMENTS}
We thank Mike Boylan-Kolchin, Marla Geha, Joss Bland-Hawthorn, Alex
Ji, Ting Li, and Mike Cooper for helpful suggestions.  We are also
indebted to Alex Ji for providing his compilation of UFD abundances,
which is used in Fig.~\ref{fig:abundances}, and to Ting Li for
supplying the model stellar population upon which
Fig.~\ref{fig:sensitivity} is based.

This publication is based upon work supported by the National Science
Foundation under grants AST-1714873 and AST-1412792.  JDS also
acknowledges a productive stay during the writing of this article at
the Kavli Institute for Theoretical Physics, which is supported in
part by the National Science Foundation under Grant No. NSF
PHY-1748958, for the program \emph{The Small-Scale Structure of
  Cold(?) Dark Matter}.  This paper would only barely have been
possible without NASA's Astrophysics Data System Bibliographic
Services.

\bibliographystyle{ar-style2}

\end{document}

%% file: dwarf_tab.tex
\begin{deluxetable}{lcccccccr}
\tablecaption{Dwarf Galaxy Data}
\tablewidth{0pt}
\tablehead{
\colhead{Dwarf} &
\colhead{$M_{{\rm V}}$} &
\colhead{$R_{1/2}$} &
\colhead{Distance} &
\colhead{$v_{\rm hel}$} &
\colhead{$\sigma$} &
\colhead{\feh} &
\colhead{$\sigma_{{\rm \feh}}$} &
\colhead{References\tablenotemark{{\rm a,b}}} \\
\colhead{} &
\colhead{} &
\colhead{(pc)} &
\colhead{(kpc)} &
\colhead{(\kms)} &
\colhead{(\kms)} &
\colhead{} &
\colhead{} &
\colhead{}
}
\startdata
        Tucana~IV & $ -3.50^{+0.28}_{-0.28}$ & $ 127^{+ 26}_{- 22}$ & $ 48.0^{+ 4.0}_{- 4.0}$ & $                              $ & $                                   $  & $                     $  & $                                                        $ &         1,1,1,-,-,-,-\\
         Sculptor & $-10.82^{+0.14}_{-0.14}$ & $ 279^{+ 16}_{- 16}$ & $ 86.0^{+ 5.0}_{- 5.0}$ & $     $\phs$111.4^{+0.1}_{-0.1}$ & $            $\phn$9.2^{+1.1}_{-1.1}$  & $-1.73^{+0.03}_{-0.02}$  & $                  0.44                  ^{+0.02}_{-0.02}$ &         2,2,3,4,5,6,6\\
         Cetus~II & $  0.00^{+0.68}_{-0.68}$ & $  17^{+  9}_{-  5}$ & $ 30.0^{+ 3.0}_{- 3.0}$ & $                              $ & $                                   $  & $                     $  & $                                                        $ &         1,1,1,-,-,-,-\\
        Cetus~III & $ -2.45^{+0.57}_{-0.56}$ & $  90^{+ 32}_{- 14}$ & $251.0^{+24.0}_{-11.0}$ & $                              $ & $                                   $  & $                     $  & $                                                        $ &         7,7,7,-,-,-,-\\
    Triangulum~II & $ -1.60^{+0.76}_{-0.76}$ & $  16^{+  4}_{-  4}$ & $ 28.4^{+ 1.6}_{- 1.6}$ & $          -381.7^{+1.1}_{-1.1}$ & $<3.4\tablenotemark{c}              $  & $-2.24^{+0.05}_{-0.05}$  & $                  0.53                  ^{+0.12}_{-0.38}$ &         2,2,8,9,9,9,9\\
          Segue~2 & $ -1.98^{+0.88}_{-0.88}$ & $  40^{+  4}_{-  4}$ & $ 37.0^{+ 3.0}_{- 3.0}$ & $     $\phn$-40.2^{+0.9}_{-0.9}$ & $<2.2\tablenotemark{c}              $  & $-2.14^{+0.16}_{-0.15}$  & $                  0.39                  ^{+0.12}_{-0.13}$ &      2,2,10,11,11,6,6\\
    DESJ0225+0304 & $ -1.10^{+0.50}_{-0.30}$ & $  19^{+  9}_{-  5}$ & $ 23.8^{+ 0.7}_{- 0.5}$ & $                              $ & $                                   $  & $                     $  & $                                                        $ &      12,12,12,-,-,-,-\\
         Hydrus~I & $ -4.71^{+0.08}_{-0.08}$ & $  53^{+  4}_{-  4}$ & $ 27.6^{+ 0.5}_{- 0.5}$ & $  $\phs\phn$80.4^{+0.6}_{-0.6}$ & $            $\phn$2.7^{+0.5}_{-0.4}$  & $-2.52^{+0.09}_{-0.09}$  & $                  0.41                  ^{+0.08}_{-0.08}$ &  13,13,13,13,13,13,13\\
           Fornax & $-13.34^{+0.14}_{-0.14}$ & $ 792^{+ 18}_{- 18}$ & $139.0^{+ 3.0}_{- 3.0}$ & $  $\phs\phn$55.2^{+0.1}_{-0.1}$ & $                 11.7^{+0.9}_{-0.9}$  & $-1.07^{+0.02}_{-0.01}$  & $                  0.27                  ^{+0.01}_{-0.01}$ &       2,14,15,4,5,6,6\\
     Horologium~I & $ -3.76^{+0.56}_{-0.56}$ & $  40^{+ 10}_{-  9}$ & $ 87.0^{+13.0}_{-11.0}$ & $     $\phs$112.8^{+2.5}_{-2.6}$ & $            $\phn$4.9^{+2.8}_{-0.9}$  & $-2.76^{+0.10}_{-0.10}$  & $                  0.17                  ^{+0.20}_{-0.03}$ &    2,2,16,17,18,18,18\\
    Horologium~II & $ -1.56^{+1.02}_{-1.02}$ & $  44^{+ 15}_{- 14}$ & $ 78.0^{+ 8.0}_{- 7.0}$ & $                              $ & $                                   $  & $                     $  & $                                                        $ &        2,2,19,-,-,-,-\\
     Reticulum~II & $ -3.99^{+0.38}_{-0.38}$ & $  51^{+  3}_{-  3}$ & $ 31.6^{+ 1.5}_{- 1.4}$ & $  $\phs\phn$62.8^{+0.5}_{-0.5}$ & $            $\phn$3.3^{+0.7}_{-0.7}$  & $-2.65^{+0.07}_{-0.07}$  & $                  0.28                  ^{+0.09}_{-0.09}$ &    2,2,20,21,21,21,21\\
      Eridanus~II & $ -7.10^{+0.30}_{-0.30}$ & $ 246^{+ 17}_{- 17}$ & $366.0^{+17.0}_{-17.0}$ & $  $\phs\phn$75.6^{+1.3}_{-1.3}$ & $            $\phn$6.9^{+1.2}_{-0.9}$  & $-2.38^{+0.13}_{-0.13}$  & $                  0.47                  ^{+0.12}_{-0.09}$ &  22,22,22,23,23,23,23\\
    Reticulum~III & $ -3.30^{+0.29}_{-0.29}$ & $  64^{+ 26}_{- 23}$ & $ 92.0^{+13.0}_{-13.0}$ & $                              $ & $                                   $  & $                     $  & $                                                        $ &         1,1,1,-,-,-,-\\
         Pictor~I & $ -3.67^{+0.60}_{-0.60}$ & $  32^{+ 15}_{- 15}$ & $126.0^{+19.0}_{-16.0}$ & $                              $ & $                                   $  & $                     $  & $                                                        $ &        2,2,16,-,-,-,-\\
        Columba~I & $ -4.20^{+0.20}_{-0.20}$ & $ 117^{+ 12}_{- 12}$ & $183.0^{+10.0}_{-10.0}$ & $                              $ & $                                   $  & $                     $  & $                                                        $ &         8,8,8,-,-,-,-\\
           Carina & $ -9.45^{+0.05}_{-0.05}$ & $ 311^{+ 15}_{- 15}$ & $106.0^{+ 5.0}_{- 5.0}$ & $     $\phs$222.9^{+0.1}_{-0.1}$ & $            $\phn$6.6^{+1.2}_{-1.2}$  & $-1.80^{+0.02}_{-0.02}$  & $                  0.24   $\tablenotemark{d}               &      2,2,24,4,5,25,25\\
        Pictor~II & $ -3.20^{+0.40}_{-0.50}$ & $  47^{+ 20}_{- 13}$ & $ 45.0^{+ 5.0}_{- 4.0}$ & $                              $ & $                                   $  & $                     $  & $                                                        $ &      26,26,26,-,-,-,-\\
        Carina~II & $ -4.50^{+0.10}_{-0.10}$ & $  92^{+  8}_{-  8}$ & $ 36.2^{+ 0.6}_{- 0.6}$ & $     $\phs$477.2^{+1.2}_{-1.2}$ & $            $\phn$3.4^{+1.2}_{-0.8}$  & $-2.44^{+0.09}_{-0.09}$  & $                  0.22                  ^{+0.10}_{-0.07}$ &  27,27,27,28,28,28,28\\
       Carina~III & $ -2.40^{+0.20}_{-0.20}$ & $  30^{+  8}_{-  8}$ & $ 27.8^{+ 0.6}_{- 0.6}$ & $     $\phs$284.6^{+3.4}_{-3.1}$ & $            $\phn$5.6^{+4.3}_{-2.1}$  & $                     $  & $                                                        $ &    27,27,27,28,28,-,-\\
    Ursa~Major~II & $ -4.43^{+0.26}_{-0.26}$ & $ 139^{+  9}_{-  9}$ & $ 34.7^{+ 2.0}_{- 1.9}$ & $          -116.5^{+1.9}_{-1.9}$ & $            $\phn$5.6^{+1.4}_{-1.4}$  & $-2.23^{+0.21}_{-0.24}$  & $                  0.67                  ^{+0.20}_{-0.15}$ &      2,2,29,30,31,6,6\\
            Leo~T & $ -8.00$\tablenotemark{e} & $ 118^{+ 11}_{- 11}$ & $409.0^{+29.0}_{-27.0}$ & $  $\phs\phn$38.1^{+2.0}_{-2.0}$ & $ $\phn$7.5^{+1.6}_{-1.6}$  & $-1.91^{+0.12}_{-0.14}$  & $ 0.43^{+0.13}_{-0.09}$ &    32,32,33,30,30,6,6\\
          Segue~1 & $ -1.30^{+0.73}_{-0.73}$ & $  24^{+  4}_{-  4}$ & $ 23.0^{+ 2.0}_{- 2.0}$ & $     $\phs$208.5^{+0.9}_{-0.9}$ & $            $\phn$3.7^{+1.4}_{-1.1}$  & $-2.71^{+0.45}_{-0.39}$  & $                  0.95                  ^{+0.42}_{-0.26}$ &    2,2,34,35,35,36,36\\
            Leo~I & $-11.78^{+0.28}_{-0.28}$ & $ 270^{+ 17}_{- 16}$ & $254.0^{+16.0}_{-15.0}$ & $     $\phs$282.9^{+0.5}_{-0.5}$ & $            $\phn$9.2^{+0.4}_{-0.4}$  & $-1.48^{+0.02}_{-0.01}$  & $                  0.26                  ^{+0.01}_{-0.01}$ &      2,2,37,38,38,6,6\\
          Sextans & $ -8.94^{+0.06}_{-0.06}$ & $ 456^{+ 15}_{- 15}$ & $ 95.0^{+ 3.0}_{- 3.0}$ & $     $\phs$224.3^{+0.1}_{-0.1}$ & $            $\phn$7.9^{+1.3}_{-1.3}$  & $-1.97^{+0.04}_{-0.04}$  & $                  0.38                  ^{+0.03}_{-0.03}$ &        2,2,39,4,5,6,6\\
     Ursa~Major~I & $ -5.13^{+0.38}_{-0.38}$ & $ 295^{+ 28}_{- 28}$ & $ 97.3^{+ 6.0}_{- 5.7}$ & $     $\phn$-55.3^{+1.4}_{-1.4}$ & $            $\phn$7.0^{+1.0}_{-1.0}$  & $-2.16^{+0.11}_{-0.13}$  & $                  0.62                  ^{+0.10}_{-0.08}$ &     2,40,41,30,31,6,6\\
        Willman~1 & $ -2.90^{+0.74}_{-0.74}$ & $  33^{+  8}_{-  8}$ & $ 45.0^{+10.0}_{-10.0}$ & $     $\phn$-14.1^{+1.0}_{-1.0}$ & $            $\phn$4.0^{+0.8}_{-0.8}$  & $-2.19^{+0.08}_{-0.08}$  & $                                                        $ &     2,2,42,43,43,43,-\\
           Leo~II & $ -9.74^{+0.04}_{-0.04}$ & $ 171^{+ 10}_{- 10}$ & $233.0^{+14.0}_{-14.0}$ & $  $\phs\phn$78.3^{+0.6}_{-0.6}$ & $            $\phn$7.4^{+0.4}_{-0.4}$  & $-1.68^{+0.02}_{-0.03}$  & $                  0.34                  ^{+0.02}_{-0.02}$ &      2,2,44,45,45,6,6\\
            Leo~V & $ -4.29^{+0.36}_{-0.36}$ & $  49^{+ 16}_{- 16}$ & $169.0^{+ 4.0}_{- 4.0}$ & $     $\phs$170.9^{+2.1}_{-1.9}$ & $            $\phn$2.3^{+3.2}_{-1.6}$  & $-2.48^{+0.21}_{-0.21}$  & $                  0.47                  ^{+0.23}_{-0.13}$ &    2,2,46,47,47,47,47\\
           Leo~IV & $ -4.99^{+0.26}_{-0.26}$ & $ 114^{+ 13}_{- 13}$ & $154.0^{+ 5.0}_{- 5.0}$ & $     $\phs$132.3^{+1.4}_{-1.4}$ & $            $\phn$3.3^{+1.7}_{-1.7}$  & $-2.29^{+0.19}_{-0.22}$  & $                  0.56                  ^{+0.19}_{-0.14}$ &      2,2,48,30,30,6,6\\
        Crater~II & $ -8.20^{+0.10}_{-0.10}$ & $1066^{+ 86}_{- 86}$ & $117.5^{+ 1.1}_{- 1.1}$ & $  $\phs\phn$87.5^{+0.4}_{-0.4}$ & $            $\phn$2.7^{+0.3}_{-0.3}$  & $-1.98^{+0.10}_{-0.10}$  & $                  0.22                  ^{+0.04}_{-0.03}$ &  49,49,49,50,50,50,50\\
          Virgo~I & $ -0.80^{+0.90}_{-0.90}$ & $  38^{+ 12}_{- 11}$ & $ 87.0^{+13.0}_{- 8.0}$ & $                              $ & $                                   $  & $                     $  & $                                                        $ &      51,51,51,-,-,-,-\\
         Hydra~II & $ -4.86^{+0.37}_{-0.37}$ & $  67^{+ 13}_{- 13}$ & $151.0^{+ 8.0}_{- 7.0}$ & $     $\phs$303.1^{+1.4}_{-1.4}$ & $<3.6\tablenotemark{c}              $  & $-2.02^{+0.08}_{-0.08}$  & $                  0.40                  ^{+0.48}_{-0.26}$ &    2,2,52,53,53,53,53\\
   Coma~Berenices & $ -4.28^{+0.25}_{-0.25}$ & $  69^{+  5}_{-  4}$ & $ 42.0^{+ 1.6}_{- 1.5}$ & $  $\phs\phn$98.1^{+0.9}_{-0.9}$ & $            $\phn$4.6^{+0.8}_{-0.8}$  & $-2.43^{+0.11}_{-0.11}$  & $                  0.46                  ^{+0.09}_{-0.08}$ &      2,2,54,30,30,6,6\\
Canes~Venatici~II & $ -5.17^{+0.32}_{-0.32}$ & $  71^{+ 11}_{- 11}$ & $160.0^{+ 4.0}_{- 4.0}$ & $          -128.9^{+1.2}_{-1.2}$ & $            $\phn$4.6^{+1.0}_{-1.0}$  & $-2.35^{+0.16}_{-0.19}$  & $                  0.57                  ^{+0.15}_{-0.12}$ &      2,2,55,30,30,6,6\\
 Canes~Venatici~I & $ -8.73^{+0.06}_{-0.06}$ & $ 437^{+ 18}_{- 18}$ & $211.0^{+ 6.0}_{- 6.0}$ & $  $\phs\phn$30.9^{+0.6}_{-0.6}$ & $            $\phn$7.6^{+0.4}_{-0.4}$  & $-1.91^{+0.04}_{-0.04}$  & $                  0.39                  ^{+0.03}_{-0.02}$ &      2,2,56,30,30,6,6\\
    Bo{\"o}tes~II & $ -2.94^{+0.74}_{-0.75}$ & $  39^{+  5}_{-  5}$ & $ 42.0^{+ 1.0}_{- 1.0}$ & $          -117.0^{+5.2}_{-5.2}$ & $                 10.5^{+7.4}_{-7.4}$  & $-2.79^{+0.06}_{-0.10}$  & $<0.35\tablenotemark{c}                                  $ &    2,2,57,58,58,59,59\\
     Bo{\"o}tes~I & $ -6.02^{+0.25}_{-0.25}$ & $ 191^{+  8}_{-  8}$ & $ 66.0^{+ 2.0}_{- 2.0}$ & $     $\phs$101.8^{+0.7}_{-0.7}$ & $            $\phn$4.6^{+0.8}_{-0.6}$  & $-2.35^{+0.09}_{-0.08}$  & $                  0.44                  ^{+0.07}_{-0.06}$ &    2,2,60,61,61,62,62\\
       Ursa~Minor & $ -9.03^{+0.05}_{-0.05}$ & $ 405^{+ 21}_{- 21}$ & $ 76.0^{+ 4.0}_{- 4.0}$ & $          -247.2^{+0.8}_{-0.8}$ & $            $\phn$9.5^{+1.2}_{-1.2}$  & $-2.12^{+0.03}_{-0.02}$  & $                  0.33                  ^{+0.02}_{-0.03}$ &       2,2,63,64,4,6,6\\
         Draco~II & $ -0.80^{+0.40}_{-1.00}$ & $  19^{+  4}_{-  3}$ & $ 21.5^{+ 0.4}_{- 0.4}$ & $          -342.5^{+1.1}_{-1.2}$ & $<5.9\tablenotemark{c}              $  & $-2.70^{+0.10}_{-0.10}$  & $<0.24\tablenotemark{c}                                  $ &  65,65,65,65,65,65,65\\
         Hercules & $ -5.83^{+0.17}_{-0.17}$ & $ 216^{+ 20}_{- 20}$ & $132.0^{+ 6.0}_{- 6.0}$ & $  $\phs\phn$45.0^{+1.1}_{-1.1}$ & $            $\phn$5.1^{+0.9}_{-0.9}$  & $-2.47^{+0.13}_{-0.12}$  & $                  0.47                  ^{+0.11}_{-0.08}$ &      2,2,66,30,30,6,6\\
            Draco & $ -8.88^{+0.05}_{-0.05}$ & $ 231^{+ 17}_{- 17}$ & $ 82.0^{+ 6.0}_{- 6.0}$ & $          -290.7^{+0.7}_{-0.8}$ & $            $\phn$9.1^{+1.2}_{-1.2}$  & $-2.00^{+0.02}_{-0.02}$  & $                  0.34                  ^{+0.02}_{-0.02}$ &       2,2,67,64,4,6,6\\
      Sagittarius & $-13.50^{+0.15}_{-0.15}$ & $2662^{+193}_{-193}$ & $ 26.7^{+ 1.3}_{- 1.3}$ & $     $\phs$139.4^{+0.6}_{-0.6}$ & $            $\phn$9.6^{+0.4}_{-0.4}$  & $-0.53^{+0.03}_{-0.02}$  & $                  0.17                  ^{+0.02}_{-0.02}$ &  68,68,69,70,70,71,71\\
   Sagittarius~II & $ -5.20^{+0.10}_{-0.10}$ & $  33^{+  2}_{-  2}$ & $ 70.1^{+ 2.3}_{- 2.3}$ & $                              $ & $                                   $  & $                     $  & $                                                        $ &      20,20,20,-,-,-,-\\
         Indus~II & $ -4.30^{+0.19}_{-0.19}$ & $ 181^{+ 70}_{- 64}$ & $214.0^{+16.0}_{-16.0}$ & $                              $ & $                                   $  & $                     $  & $                                                        $ &         1,1,1,-,-,-,-\\
          Grus~II & $ -3.90^{+0.22}_{-0.22}$ & $  93^{+ 16}_{- 12}$ & $ 53.0^{+ 5.0}_{- 5.0}$ & $                              $ & $                                   $  & $                     $  & $                                                        $ &         1,1,1,-,-,-,-\\
      Pegasus~III & $ -4.10^{+0.50}_{-0.50}$ & $  78^{+ 31}_{- 25}$ & $205.0^{+20.0}_{-20.0}$ & $          -222.9^{+2.6}_{-2.6}$ & $            $\phn$5.4^{+3.0}_{-2.5}$  & $-2.40^{+0.15}_{-0.15}$  & $                                                        $ &   72,72,72,73,73,73,-\\
      Aquarius~II & $ -4.36^{+0.14}_{-0.14}$ & $ 160^{+ 26}_{- 26}$ & $107.9^{+ 3.3}_{- 3.3}$ & $     $\phn$-71.1^{+2.5}_{-2.5}$ & $            $\phn$5.4^{+3.4}_{-0.9}$  & $-2.30^{+0.50}_{-0.50}$  & $                                                        $ &   74,74,74,74,74,49,-\\
        Tucana~II & $ -3.90^{+0.20}_{-0.20}$ & $ 121^{+ 35}_{- 35}$ & $ 58.0^{+ 8.0}_{- 8.0}$ & $          -129.1^{+3.5}_{-3.5}$ & $            $\phn$8.6^{+4.4}_{-2.7}$  & $-2.90^{+0.15}_{-0.16}$  & $                  0.29                  ^{+0.15}_{-0.12}$ &  16,16,16,75,75,76,76\\
           Grus~I & $ -3.47^{+0.59}_{-0.59}$ & $  28^{+ 23}_{- 23}$ & $120.0^{+12.0}_{-11.0}$ & $          -140.5^{+2.4}_{-1.6}$ & $            $\phn$2.9^{+2.1}_{-1.0}$  & $-1.42^{+0.55}_{-0.42}$  & $                  0.41                  ^{+0.49}_{-0.23}$ &    2,2,17,75,75,75,75\\
        Pisces~II & $ -4.23^{+0.38}_{-0.38}$ & $  60^{+ 10}_{- 10}$ & $183.0^{+15.0}_{-15.0}$ & $          -226.5^{+2.7}_{-2.7}$ & $            $\phn$5.4^{+3.6}_{-2.4}$  & $-2.45^{+0.07}_{-0.07}$  & $                  0.48                  ^{+0.70}_{-0.29}$ &    2,2,77,53,53,53,53\\
         Tucana~V & $ -1.60^{+0.49}_{-0.49}$ & $  16^{+  5}_{-  5}$ & $ 55.0^{+ 9.0}_{- 9.0}$ & $                              $ & $                                   $  & $                     $  & $                                                        $ &         1,1,1,-,-,-,-\\
       Phoenix~II & $ -2.70^{+0.40}_{-0.40}$ & $  37^{+  8}_{-  8}$ & $ 84.3^{+ 4.0}_{- 4.0}$ & $                              $ & $                                   $  & $                     $  & $                                                        $ &      20,20,20,-,-,-,-\\
       Tucana~III & $ -1.49^{+0.20}_{-0.20}$ & $  37^{+  9}_{-  9}$ & $ 25.0^{+ 2.0}_{- 2.0}$ & $          -102.3^{+0.4}_{-0.4}$ & $<1.2\tablenotemark{c}              $  & $-2.42^{+0.07}_{-0.08}$  & $<0.19\tablenotemark{c}                                  $ &   20,20,1,78,78,78,78\\
\enddata
\vspace{-0.3in}
\tablecomments{These data are provided as a convenience to the community.  However, in recognition of the effort invested by many researchers to obtain, reduce, analyze, and publish these measurements, we strongly encourage authors to cite the original references (which are listed below), not just this compilation, where possible.}
\tablenotetext{a}{References:  (1)  \citet{drlica15};  (2)  \citet{munoz18b};  (3)  \citet{pietrzynski08};  (4)  \citet{walker09};  (5)  \citet{walker09c};  (6)  \citet{kirby13b};  (7)  \citet{homma18};  (8)  \citet{carlin17};  (9)  \citet{kirby17};  (10)  \citet{boettcher13};  (11)  \citet{kirby13};  (12)  \citet{luque17};  (13)  \citet{koposov18};  (14)  \citet{battaglia06};  (15)  \citet{rizzi07};  (16)  \citet{bechtol15};  (17)  \citet{koposov15};  (18)  \citet{koposov15b};  (19)  \citet{kj15};  (20)  \citet{mutlu18};  (21)  \citet{simon15};  (22)  \citet{crnojevic16};  (23)  \citet{li17};  (24)  \citet{karczmarek15};  (25)  \citet{fabrizio12};  (26)  \citet{dw16};  (27)  \citet{torrealba18};  (28)  \citet{li18};  (29)  \citet{dallora12};  (30)  \citet{sg07};  (31) this work (32)  \citet{dejong08};  (33)  \citet{clementini12};  (34)  \citet{belokurov07};  (35)  \citet{simon11};  (36)  \citet{fsk14};  (37)  \citet{bellazzini04};  (38)  \citet{mateo08};  (39)  \citet{lee03};  (40)  \citet{okamoto08};  (41)  \citet{garofalo13};  (42)  \citet{willman05a};  (43)  \citet{willman11};  (44)  \citet{bellazzini05};  (45)  \citet{spencer17};  (46)  \citet{medina18};  (47)  \citet{collins17};  (48)  \citet{moretti09};  (49)  \citet{torrealba16};  (50)  \citet{caldwell17};  (51)  \citet{homma16};  (52)  \citet{vivas16};  (53)  \citet{kirby15};  (54)  \citet{musella09};  (55)  \citet{greco08};  (56)  \citet{kuehn08};  (57)  \citet{walsh08};  (58)  \citet{koch09};  (59)  \citet{ji16b};  (60)  \citet{dallora06};  (61)  \citet{koposov11};  (62)  \citet{brown14};  (63)  \citet{bellazzini02};  (64)  \citet{munoz05};  (65)  \citet{longeard18};  (66)  \citet{musella12};  (67)  \citet{kinemuchi08};  (68)  \citet{majewski03};  (69)  \citet{hamanowicz16};  (70)  \citet{bellazzini08};  (71)  \citet{mucciarelli17};  (72)  \citet{kim15peg3};  (73)  \citet{kim16};  (74)  \citet{torrealba16b};  (75)  \citet{walker16};  (76)  \citet{chiti18};  (77)  \citet{sand12};  (78)  \citet{simon17}.}
\tablenotetext{b}{The references listed for each object are for, in order: (1) $M_{V}$, (2) $R_{1/2}$, (3) distance, (4) $v_{{\rm hel}}$, (5) $\sigma$, (6) [Fe/H], and (7) $\sigma_{{\rm \feh}}$.  Inasmuch as the properties of some galaxies have been determined by multiple studies, this reference list is not intended to be complete.  Instead, it represents our assessment of the best available data.  In cases where no velocity and/or metallicity measurements are available in the literature, a dash is listed in place of the corresponding reference.}
\tablenotetext{c}{Upper limits are at 90\% confidence.  Where the original reference does not provide a value at that confidence interval, we have determined one from the data.}
\tablenotetext{d}{No uncertainty on the metallicity dispersion of Carina was provided by \citet{fabrizio12}.}
\tablenotetext{e}{No uncertainty on the absolute magnitude of Leo~T was provided by \citet{dejong08}.}
\label{data_table}
\end{deluxetable}